\documentclass[aps,prd,preprint]{revtex4}
\usepackage{graphicx}
\usepackage{longtable}

\def\dmu{\Delta\mu}
\def\be{\begin{equation}}
\def\ee{\end{equation}}

\begin{document}

\title{The galaxy ancestor problem}

\author{M.J. Disney}
\email{mjdisney@gmail.com}
\affiliation{School of Physics and Astronomy, Cardiff University, The Parade, Cardiff, CF24 3AA, Wales, UK}

\author{R.H. Lang}
\affiliation{School of Physics and Astronomy, Cardiff University, The Parade, Cardiff, CF24 3AA, Wales, UK}

\date{\today}




\begin{abstract}
HST finds galaxies whose Tolman dimming should exceed 10 mag. Could
evolution alone explain these as our ancestor galaxies? Or could they
be representatives of quite a different dynasty whose descendents are
no longer prominent today? We explore this latter hypothesis and argue
that Surface Brightness Selection Effects naturally bring into focus
quite different dynasties from different redshifts. Thus the HST $z=7$
galaxies could be examples of galaxies whose descendents are both too
small and too choked with dust to be recognisable in our neighbourhood
easily today. Conversely the ancestors of the Milky Way and its
obvious neighbours will have completely sunk below the sky at $z > 1.2$,
although their diffused light could account for the missing
Reionization flux. This Succeeding Prominent Dynasties Hypothesis
(SPDH) fits the existing observations both naturally and well,
including the bizarre distributions of galaxy surface
brightnessâ found in deep fields, the angular size $\sim (1+z)^{-1}$ law,
'downsizing'â which turns out to be an
 'illusion'â in the sense that it is does not imply evolution,
'Infant Mortality', i.e. the discrepancy between stars born
and stars seen, and finally the recently discovered and unexpected
excess of QSOAL DLAs at high redshift. If the SPDH is true then a
large proportion of galaxies remain sunk from sight, probably at all
redshifts. We show that fishing them out of the sky by their optical
emissions alone will be practically impossible, even when they are
nearby. More ingenious methods will be needed to detect them. It
follows that disentangling galaxy evolution through studying ever
higher redshift galaxies may be a forlorn hope because one will be
comparing young oranges with old apples, not ancestors with their true
descendants.
\end{abstract}


\maketitle

\section{Introduction}

Attempts to decipher the evolution of the cosmos through studying
high-redshift galaxies rely on the implicit assumption that those
galaxies are, in some sense, the ancestors of the galaxies around us
today. But what if they are not? We would not be comparing like with
like and so be completely misled.

Tolman (1930) long ago argued that the surface brightnesses of
galaxies would dim with redshift z as $(1+z)^{-4}$ , indeed proposed it
as a test for expansion. Now that the new wide-field camera WFC-3 on
Hubble( Mackenty et al 2010) can routinely find galaxies at redshifts
of 7 or more this raises serious questions as to their nature. Their
Surface Ð Brightnesses (SBs) as measured in our frame are 
to those of galaxies nearby, such as the Milky
Way. Thus to be the ancestors of the local population they must have
undergone enormous evolution (dimming by ~$\sim$9 magnitudes) Ð in
lockstep with redshift. This might seem a fortuitous
coincidence,  particularly when the star formation histories of
local galaxies show few signs of such dramatic evolution ,
testifying more to fairly constant rates of star formation throughout
cosmic time, e.g. Tosi (2008).

Here we explore an alternative hypothesis: that the populations of
galaxies which will show up at different redshifts are different from
one another. They are not ancestors and descendants, but members of
quite distinct families. For instance galaxies prominent at high
redshift may be a physically compact, very high SB family which can
take a lot of Tolman dimming, ( $\geq$ 10 magnitudes,) without
disappearing from our sight at redshift 7 or more. The problem then
becomes explaining where their descendants are today. The HST
observations show that they are very small (sub kpc),  dense, 
rather rare in co-moving density terms, and have no dust
absorption. Taking into account their small sizes, and self-absorption
by dust, which would naturally be high in such systems today, their
contemporary descendants might be inconspicuous amongst the population
of currently prominent galaxies.

Conversely, as we shall show, lacking dramatic evolution, more than
half the light from a Milky Way will appear to have sunk beneath the
sky at redshift 0.5, and every last photon by redshift 1.2. Our
predecessor galaxies might therefore be totally invisible as
individuals at higher redshifts, although their integrated light could
very well swamp the output of those few compact high-z galaxies we can
still detect out there. It hardly needs to be said that such a
population of ÔSunken GalaxiesÕ could dramatically impact our
ideas of cosmic evolution. For instance they could supply the
presently missing ultraviolet photons needed to re-ionise the
Universe. They could also explain the excess of QSOALs, while
Lilly-Madau plots showing the combined star-formation rates in the
cosmos as a function of redshift would have to be seriously modified.

This hypothesis of ÔSucceeding Prominent DynastiesÕ
(ÔSPDHÓ), as opposed to the current notion of an ÔEvolving
Single Dynasty HypothesisÕ (ESDH) has its roots in a number of
older ideas. It is forgotten today, but before the Hubble was launched
it was anticipated that Tolman dimming would rob the sky of almost all
high-z galaxies, and it should have come as more of a surprise to find
that this was not the case. Local
galaxies tend to have a rather tight distribution of Surface
BrightnessÕs, the explanation for which is still controversial
(e.g. Davies, Impey \& Phillipps 1999). But if it is a
selection-effect the families of the ÔwrongÕ SB at any redshift
will appear inconspicuous by comparison with other families of the
rightÕ SB. An observer looking back through redshift space would
thus expect to see, thanks to Tolman, different prominent families at
different epochs. In particular he or she would expect to see the more
compact objects at higher redshifts, and would find angular diameters
$\propto (1+z)^{-1}$ , which is exactly observed to be the case (Sect VI).

Some of these ideas were explored in 'The Visibility of High
Redshift Galaxies' ( Phillipps, Davies \& Disney 1990) which built
on earlier papers in 1983 (Disney \& Phillipps) and 1976
(Disney). However the highest redshift being considered there and then
was 0.3! The situation has certainly moved on in a number of respects;
the observations of course, the supercession of photography by linear
electronic detectors Ð which makes the analysis markedly simpler,
and the most fashionable cosmological model in which to set the
calculations. Most importantly though, those earlier papers were
missing a vital argument about the way to normalise Visibility, an
argument that is here supplied in Sect 4, and which makes a
significant difference to the main inferences.

The purpose of this paper is to push the ÔSucceeding Prominent
Dynasties hypothesis (ÔSPDHÕ ) to the highest redshifts
currently accessible to observation ($\sim$ 10). If it can be tested
to destruction so much the better because, if it is true, then
deciphering galaxy evolution will be very much harder, and perhaps
impossible for generations to come.

 The rest of the paper is arranged by section as follows:
 
 (II) "The Narrow Window" gives a schematic outline of how the
 hypothesis works, and some of the conclusions it leads to.
 
 (III) "Galaxy Visibility Theory" demonstrates by calculation the
 non-intuitive but dramatic nature of surface brightness
 selection effects, i.e. how two plunging curves mean that only
 galaxies huddled perilously close to the sky will be seen to any
 great distance.
 
 (IV)"Imprisoned by Light" introduces a vital new argument to
 normalise Galaxy Visibility. It leads to the daunting conclusion that
 Low surface brightness galaxies too dim to turn up in the Schmidt
 photographic surveys will never be detectable in the optical, at
 least not for generations to come.Õ Thus whole dynasties of sunken
 galaxies could exist, lurking just beneath the sky.
 
 (V) "How galaxies sink from sight" incorporates Tolman dimming
 and cosmology into Visibility Theory to show how quickly redshift can
 drag galaxies below the sky. Thus Milky Ways would appear half sunk
 by redshift 0.5 and wholly sunk by z =1.2
 
 (VI)"Why high redshift galaxies look small" argues that a
 combination of high intrinsic SB and Aberration will, at high z,
 bring to the surface an extremely compact dynasty of galaxies that
 are relatively inconspicuous nearby. Their apparent angular sizes
 will obey the angular diameter $\sim (1+z)^{-1}$ law Ð as
 observed.

(VII) "The Descendants Problem" explains why the aforementioned
 z$\sim7$ galaxies can leave descendants in our neighbourhood which we
 wouldnÕt find without a dedicated search partly because they will have choked on their own dust.

(VIII)"How Ellipticals sink". repeats the Visibility Theory of
 Section III but for giant Ellipticals which have a different light distribution. They should sink more slowly
 with redshift, leading to the illusion that they formed earlier than
 spirals. Fig 9 demonstrates how perilously close all visibly
 prominent galaxies must huddle to the sky.
 
 (IX)"Downsizing, a different explanation." argues that because
 low SB galaxies sink at lower redshifts, there will be a downsizing
 illusionÕ which has nothing to do with evolution but reflects a
 correlation between intrinsic SB and luminosity in the sense that intrinsically
 less luminous galaxies generally have dimmer intrinsic surface
 brightnesses. We briefly speculate about the so called ÔMissing
 DwarfsÕ predicted by CDM.
 
     (X) The "Discussion" covers several phenomena
 predicted by the SPDH including : (a) "Infant Mortality" Ð the mismatch
 between the number of galaxies seen forming and the number later on
 seen. This is rather direct evidence that most high redshift galaxies
 have indeed sunk. (b) "Unexpected QSOALs" Ð the surprising
 number of DLAs recently found at high redshift ; more evidence of a
 sunken population, and (c) "Reionization" Ð which can be
 explained by the diffused light of all the sunken dynasties . We
 conclude that the SPDH fits the high redshift galaxy observations
 in a natural and parsimonious way. It remains to be tested by looking
 for the Sunken and Choked Galaxies predicted to lie in large numbers,
 both near and far.
 
 \section{THE NARROW WINDOW}
 
As anyone who has looked for M31 can testify, the problem of detecting galaxies in the optical is not so much lack
of light as lack of contrast against the foreground sky[M31 has a V mag of 3.4 which spread over its size of roughly 3 by 1 degrees amounts to a SB = 21.2 V mag per sq arc sec , where the sky is about 21.5 at a fair site] This can be
quantified by calculating the Visibility V of a galaxy as a function
of both its Luminosity L and its Surface-Brightness-Contrast with the
 sky (in magnitudes per square arc second or $\mu$), where the
Visibility V is the relative volume in which it could be detected. The
result is shown schematically in Fig 1. Irrespective of Luminosity
there is a very narrow window in SB contrast in which it is easy to
see galaxies. Calculations show that the FWHM for the Visibility of
Spirals and other exponential galaxies is less than 3 magnitudes. And
as is well known (Disney and Phillipps 1983, Davies et al. 1994) catalogues of galaxies
appear to conform rather well to this theoretically predicted
selection effect, though how many Ôhidden galaxiesÕ lie
undetected outside this narrow Visibility window is still a large and
open question (Impey and Bothun 1997, Davies, Impey and Phillipps
1999) Some certainly exist: on the low SB side lie Local Group
galaxies discovered as a result of enhanced star-counts, objects like
Segue 1 with SBs $\sim$ 6 mag dimmer than the peak in Fig 1 (Belokorov et
al 2007); and on the high SB side Ultra-Compact Dwarfs distinguished
from stars by their spectroscopic signatures with SBs $\sim$ 7 mag
higher (Phillipps et al 1998).  Astronomers are surprised to find how
narrow the SB window is. In Section III we justify it by
calculation. Here we attempt a schematic explanation.

To get into a given galaxy catalogue an object must obey two
independent criteria. It must be bright enough to be detected Ð
i.e. exceed some limiting catalogue apparent magnitude $m_c$ , yet large
enough in angular size to be detected as an extended object. That is
to say its apparent angular diameter $\theta$  , measured at some specified
isophote $\mu_c$, must exceed the minimum catalogue limit $\theta_c$ .
 
 If each galaxy is characterised by two parameters, absolute
 luminosity L, and intrinsic surface brightness ( say central surface
 brightness $\mu_0$ in mag arc $sec^{-2}$ , or effective SB $\emph{at}$ half light  $\mu_{1/2}$ ) then one can
 calculate the maximum distance $d^m$ at which it can lie and still obey the
 magnitude criterion, and  $d^\theta$, the maximum distance at which it can lie
 and yet obey the angular criterion. Both distances scale as $L^{1/2}$ so we can
 set aside Luminosity as a simple scaling parameter and investigate
 the more interesting dependencies of $d^m$ and $d^\theta$ on the surface brightness
 contrast $\dmu=\mu_c -\mu_0$ between galaxy and sky.
 
 Fig 1 illustrates what happens for objects with an exponentially
 declining light distribution ( virtually all galaxies bar Giant
 Ellipticals; see later). The dashed (green) line shows $d^m$ cubed (Volume
 not distance is the important measure), the smooth (red)  line $d^\theta$ cubed, both as a
 function of $\dmu$ , the SB contrast. High SB galaxies with large $\dmu$ s lie to the
 left, low SB ones with a small  $\dmu$ s to the right.
 
          \begin{figure*}
\begin{center}
\includegraphics[width=3in]{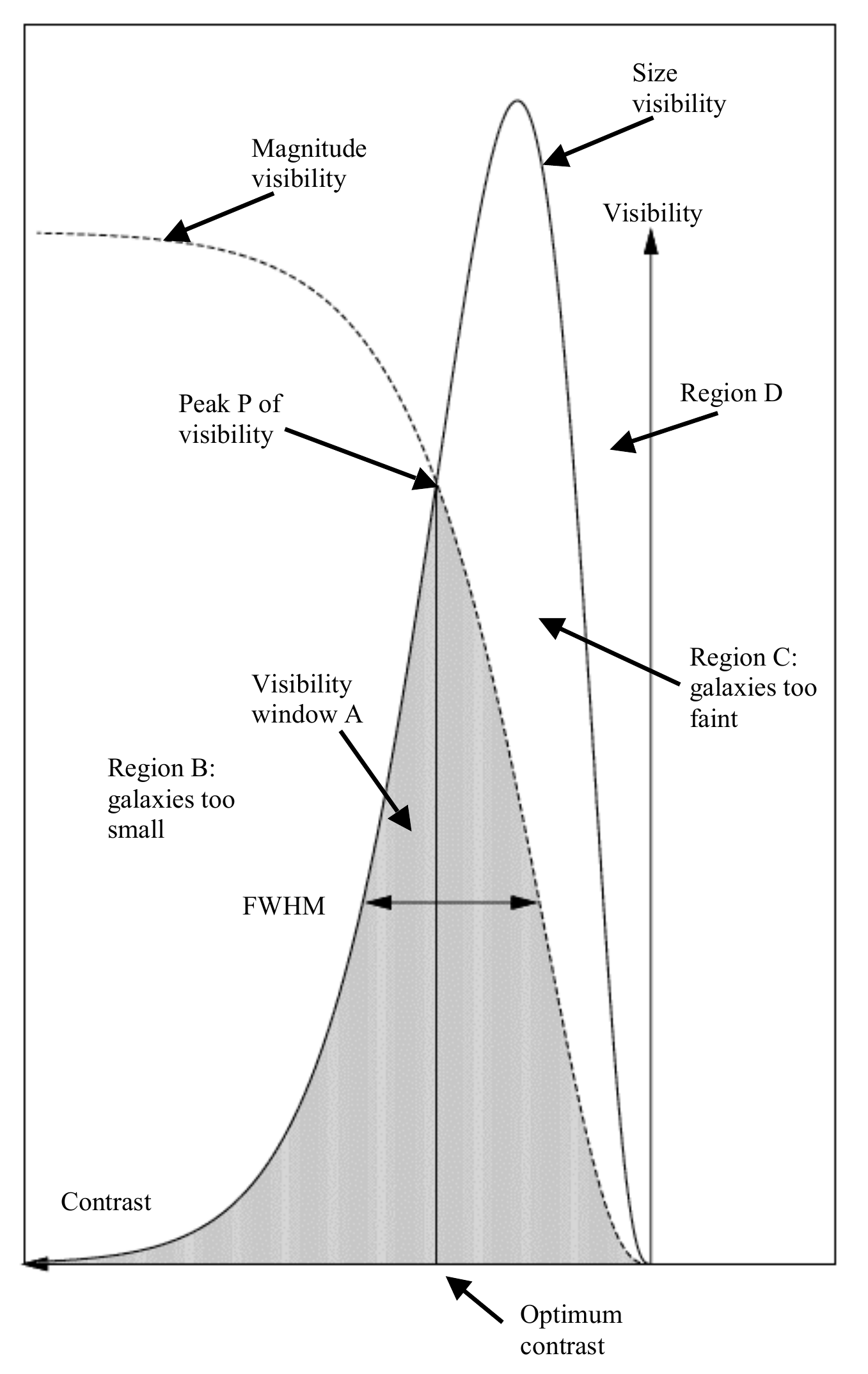}
\caption{ Schematic Wigwam diagram of the Visibility (i.e. relative
  volume in which it can be detected) of an Exponential galaxy as a
  function of its surface-brightness contrast $\dmu(mag arc sec^{-2}
  )$ to $\mu_c$ the lowest surface brightness isophote that can be
  detected in the particular survey. In the usual convention lower SBs
  are to the right, while the contrasts $\dmu$ increase to the
  left. The diagram is the same for all Luminosities Ð which only
  effect the vertical scale. The green (dashed) line is the upper
  limit to the Visibility set by the apparent magnitude limit $m_c$ of
  the survey, and so is called $V^m$ in the text. The red (smooth) line is the
  upper limit to the Visibility set by $\theta_c$ , the angular-size
  limit of the survey defined at $\mu_c$ , and so is called
  $V^{\theta}$ in the text and labelled so in Fig 3 . To be visible any galaxy must lie beneath both lines,
  and so must lie in the shaded region A. Those at the left in region
  B will be high SB objects that appear too small. Those in C will be
  low SB objects that appear too faint. Those in region D will have no
  part of their images showing above the sky; they are entirely sunk
  beneath it. In practice the FWHM of the Visible Window A is only 2.5
  magnitudes. This should be compared with Tolman dimming of 3 mag at
  a redshift of 1, and 9 mag at a redshift of 7. As one looks to
  higher redshifts so Tolman dimming will cause galaxies to march from
  left to right across the diagram, passing through the Visibility
  Window A, the Wigwam, Ð which is anchored in local coordinates by
  the brightness of the local sky (to which $\mu_c$ is related). In
  this rendering the maximum heights of the two curves dashed, and
  smooth , have been arbitrarily set roughly equal; in practice
  they can be altered by the survey parameters $m_c,\theta_c,$
  and  $\mu_c$ (See Section IV and Fig 3 for an exact rendition with all the numbers put in.)}
\label{fig1}
\end{center}
\end{figure*}   

What is going on? Consider the dashed(green) line. At high enough SB
virtually all of a galaxy's light will lie above $\mu_c$ , and $V^m \equiv (d^m)^3$ will not vary with the contrast, so the line is flat.  But as the galaxy's
SB is lowered (i.e. moves right) so the contrast $\dmu=\mu_c -\mu_0 $ drops, and more and more of its light falls below the limiting isophote $\mu_c$ , until, when $\mu_0 =\mu_c$ ( ie $\dmu=0)$ it vanishes altogether (i.e.  $d^m$ and hence $V^m \rightarrow 0 $)
 
 The red (smooth) line corresponding to $ V^\theta \equiv(d^\theta)^3$ is more interesting. It has a fairly narrow peak because at high SB ( to the left) the galaxy must
 be physically small, while to the right most of its light is
 dimmed below the limiting isophote $\mu_c$ , and what is left to measure
 above has a smaller and smaller apparent angular size until it
 vanishes altogether when $\mu_0 \rightarrow\mu_c$ and  $\dmu \rightarrow0$.  
 
 Every galaxy in the catalogue must
 obey both criteria. Thus it must lie in the hatched, Wigwam-shaped
 area A beneath both the smooth (red) line and the dashed(green) line. Both
 lines plunge steeply, resulting in a narrow FWHM with a peak at P
 corresponding to an optimum contrast $\dmu(P)$ . Higher SB galaxies in region B
 lie above the smooth (red) line, and will be too small in diameter to
 be seen as galaxies at any significant distance, while lower SB
 galaxies to the right in region C lie above the dashed (green) line
 and will be too faint to see above the sky at any greater
 distance. Galaxies in D are completely submerged below the sky, even
 their cores being dimmer than the limiting isophote $\mu_c$.
 
 Fig.1, the Visibility or Wigwam Diagram, is central to our
 hypothesis, and fundamental to galaxy research, and as such deserves
 careful study. Note first that$ \emph { is fixed in the observerÕs
 coordinate system}$ and is independent of redshift. Any galaxy that is
 redshifted, and consequently dimmed by Tolman effects, will be moved
 rightward to lower SB. A prominent or high Visibility galaxy near
 the peak at P will slide rapidly down the dashed line to the right of
 A until it is only visible nearby( Actually it will slide much faster
 because its apparent luminosity, which normalizes the height of the
 curves, is also falling at the same time due to Tolman). Note second that the
 diagram applies to all (Exponential ) galaxies, irrespective of
 Luminosity, which only changes the vertical scale. Note third
 that $\mu_c$, the outer isophotal level, will be related to the sky-brightness
 (at the appropriate wavelength) but will generally be deeper thanks
 to the accumulation of photons per detector-pixel (See Sect. IV). Fourth the HWHM of
 the Visibility Window A is generally less than 2 mags. But redshift
 dimming by $(1+z)^{-4}$ corresponds in magnitudes to +10 log(1+z) thus 2
 mags. corresponds to a redshift of less than 0.6. This implies that
 even at redshifts of a half, ancestor galaxies will be severely
 dimmed, and in many cases will be sunk out of sight entirely. So even
 at moderate redshifts ( $0.5 \leq z  \leq 1)$ the argument has to be made 
 that the galaxies we do detect out there really are the ancestors of
 the Milky Way and its catalogued neighbours.
 
 If they are not our ancestors, then what else could they be? To answer
 that it is necessary to discuss Tolman dimming. One factor of $(1+z)$
 arises from relative time-dilatation in the source, one from
 photon-weakening, i.e. photons shifting to lower energy along their
 line of flight. The other two arise from simple aberration, that is
 to say that the source was closer to the observer and therefore
 looked bigger by a factor $(1+z)$ in each dimension than it would do
 today.( i.e. the convergence angle of its light was set at emission
 not detection.)
 
Returning to Fig 1 aberration means that a source that is in region B,
and is therefore too compact to have much Visibility nearby, can be
apparently expanded by aberration and so appear relatively prominent
at higher redshifts. To understand this, note that Fig1 has no
vertical scale marked in; it shows the $ \emph{relative} $ Visibilities of
galaxies with different SB contrasts. Remove the high Visibility
galaxies (e.g. Milky Ways) by redshift-dimming then other,
intrinsically higher SB objects, will fill the peak of the Visibility
Window instead. It is always the galaxies whose apparent SBs match at
the peak (approximately at $\dmu$= 3 to 4 mags) which at any redshift will
appear most prominent, i.e. those for which 

\be \dmu' \equiv \dmu({\rm
  intrinsic})-10{\rm log}(1+z)=\dmu (P)=3.5  \ee 
  
  The narrowness of the
Visibility Window (FWHM ~$\sim$2.5 mag, as we shall prove in Sect III), by
comparison with Tolman dimming, can lead to some very surprising
phenomena and illusions. For instance:
 
(a) The apparent distribution of SBs among galaxies cannot change with
redshift, for it is a consequence of the local window. This surprising
prediction is observed( e.g. Jones and Disney 1997, see Fig
2). Tolman dimming is 3 mag by redshift 1 and 9 mag by redshift 7 ,
thus the observed constancy in Fig 2 is most unlikely to be a consequence of
dramatic stellar evolution Ð which is nowhere apparent in the
archaeology of our own and neighbouring galaxies. (e.g. Tosi 2008,
Tolstoy et al 2009)

\begin{figure*}
\begin{center}
\includegraphics[width=3in]{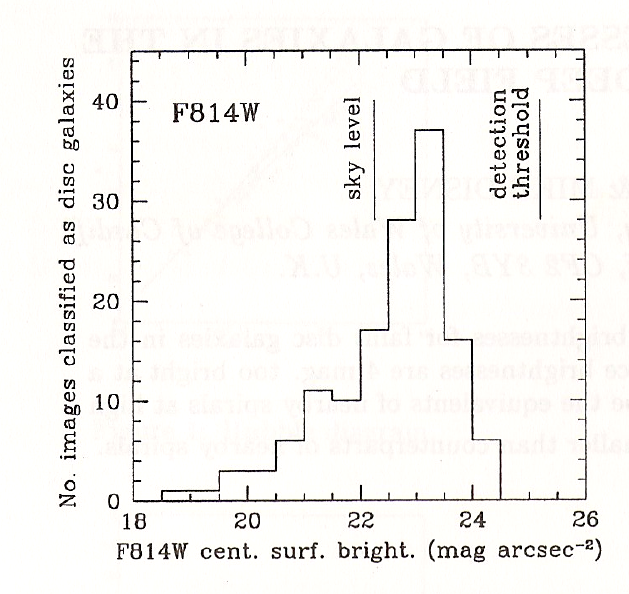}
\caption{ The distribution of central Surface brightnesses of
  exponential galaxies in a typical Hubble Deep Field Ð in this
  case the WFPC2 I-band. A connected pixel algorithm was used to
  identify images having $\geq 8$ contiguous pixels (equivalent radius
  =0.064 arc sec) above a detection threshold $\mu_L$ = 25.22 (Vega
  System) in the F814 filter. Visual morphological classification was
  performed on all images brighter than =28.0. The galaxies classified
  as exponentials, based on the presence of discs and/or their light
  profiles on visual inspection, were fitted with exponential profiles
  and hence central surface brightnesses $\mu_0's$ .(Taken from Jones
  and Disney 1997). They are fairly sharply peaked at a surface
  brightness $\mu_0$ 1 to 1.5 mag dimmer than the sky, exactly as
  predicted by Visibility Theory ( see Sect 6). Since
  such frames contain galaxies from a wide range of redshifts, and
  thus Tolman dimmings, it is very hard to understand such a sharp
  peak as anything but a profound selection effect operating in the
  observerÕs frame of reference.}\label{fig2}
\end{center}
\end{figure*}   
  
(b) Galaxies at redshifts $ > 1$ will sink below the sky, but their diffused
radiation could still dominate the universe and lead to phenomena such
as Reionisation.
  
(c) To be detected above our sky high-z galaxies must have very high
intrinsic SBs , and thus be very small for their Luminosities. Unless
galaxies are also undergoing dramatic size-evolution we must therefore
be seeing out there a new and distinctly different dynasty.
  
(d) If less luminous galaxies also have dimmer intrinsic SBs, as evidence
suggests, then that alone would lead to the illusion of
downsizingÕ i.e. dwarf galaxies will apparently only lift
themselves above the sky at recent epochs (Sect IX).
  
(e) There will be another illusion which we dub ÔInfant
MortalityÕ. Infant galaxies may briefly lift themselves above the
sky while undergoing the vigorous star-formation associated with their
birth Ð then sink from sight leaving a shortage of older
children.(Sect X)
  
(f) Those Ôdisappeared childrenÕ should nevertheless turn up in
absorption as an excess of QSOALs at high z (Sect X).
  
Once one knows what to look for, phenomena (a) to (f) are all plain to
see in the observational literature.
 
Like Anthropologists galaxy astronomers certainly have an ÔAncestor
problemÕ . However its solution may be naturally found within the
Succeeding Prominent Dynasties (SPDH) scenario .
 \section{AN OUTLINE OF VISIBILITY THEORY}
 
The Visibility of galaxies is a subtle matter with a tangled history
which, in the past, was complicated by the need to take account of
photographic saturation, no longer generally necessary. Some of
the papers were incomplete (Disney 1976, Disney and Phillipps 1983,
van der Kruit 1987) some were misleading (McGaugh 1996) and some were
wrong (Allen and Shu 1979)
 
All we attempt to do in this section, and in the simplest possible
way, is justify the narrowness of the Visibility Window A illustrated
in Fig 1 because it is so crucial to the main argument and because it
comes as such a surprise to most astronomers. To keep things simple we consider only exponential galaxies
and ignore Tolman dimming and cosmology for now (see later). If we
adopt de Vaucouleurs (1959) 2-parameter intensity I(r) profiles for
galaxies, i.e.  \be
\ln\frac{I(\theta)}{I(0)}=-\left(\frac{\theta}{\alpha}\right)^\frac{1}{\beta} \ee ($\beta$ =1 for pure
Exponentials, $\beta$=4 for Giant Ellipticals, with hybrid galaxies in between) then we can reach the main
results analytically. It is easily shown that the apparent luminosity,
integrated over the image out to angle $\Theta$ is [$\beta$ =1 henceforth until we reach Sect VIII.] \be l(\Theta)=\int_0^\Theta
2\pi \theta I(\theta) \cdot d\theta=2\pi I_{0}
\alpha^{2}\left[1-(1+\frac{\Theta}{\alpha})\cdot\exp(\frac{-\Theta}{\alpha})\right]
\ee  so that as $\Theta\rightarrow \infty$ the total apparent luminosity \be
l_T=2\pi I_0\alpha^2 \ee where  $I_0$ is the central SB and $\alpha$ the $\emph {angular}$
scale-length.  Thus (4):

\be\alpha=\frac{1}{\sqrt{2\pi}}\times\sqrt{\frac{l_T}{I_0}} \ee
 
If the angular  radius out to the outermost detectable isophote $I_c$ is $  \Theta_{out}\equiv N\times\alpha$ , which
defines N, then the perceived angular diameter $\theta=2\Theta_{out}=2N\alpha$.

Thus \be l(\Theta_{out}) =l_T \cdot [1-(1+N)\cdot exp(-N)]\ee

 From (2)
 
 \be
 \frac{\Theta_{out}}{\alpha}\equiv N=\ln (\frac{I_0}{I_c})=(0.4\ln 10)\times (\mu_c-\mu_0)
\ee

where $\mu_c$  and $\mu_0$ are $I_c $ and $I_0$ in magnitudes. So defining the $\emph{vital}$ SB contrast:
\be
\dmu\equiv(\mu_c-\mu_0)
\ee
\be
N=(0.4\ln 10)\dmu=0.92\dmu
\ee
Combining (4) (5) and (7) and recalling that
 $l=dex(-0.2m)$    and  $ I_0=dex(-0.4\mu_0)$
 
 \be
 \theta"(diam)=2N\frac{1}{\sqrt2\pi} dex[-0.2(m-\mu_0)]
 \ee

Or using  $m-M=5logd(pc)-5$
\be \theta"(diam)=\frac{10}{d(pc)}\cdot \sqrt{\frac{2}{\pi}} \cdot N \cdot dex(0.2\mu_0) \cdot dex (0.2M)\ee

And replacing $\mu_0 $  by $ \dmu$  using (8)
      
      \begin{eqnarray}
      \theta"(diam)=\frac{10}{d(pc)}\cdot \sqrt\frac{2}{\pi} \cdot \{0.92\dmu\cdot dex(-0.2\dmu)\}\nonumber\\ \times dex(0.2\mu_c)\cdot dex(-0.2M)
      \end{eqnarray}

It shows that angular size is a separable function of the absolute
magnitude M and the SB contrast $\dmu$, as one might have expected.
   
To get into a sample or catalogue with a minimum angular size $\theta_c$ a galaxy
must then be at a distance  $d^\theta$(in pc) such that : \begin{eqnarray}
 d^\theta
\leq\frac{10}{\theta_c"}.\sqrt\frac{2}{\pi}\times\{0.92\dmu.dex(-0.2\dmu)\}\nonumber\\  \times dex(0.2\mu_c).dex(-0.2M)
\end{eqnarray} which could also be written:

 \begin{eqnarray} d^\theta\leq 10\cdot
\sqrt{\frac{2}{\pi}} \times \left\{ 0.92\dmu \cdot
dex(-0.2\dmu)\right\}\nonumber\\ \times 
\left[\frac{1}{I_c\theta_c^2}\right]^{0.5}\times dex(-0.2M) \end{eqnarray} Which
neatly separates the contrast, inside curly brackets, the catalogue, inside square brackets, and the Luminosity factors in the expression for $ V^{\theta} \propto (d^\theta)^3$  . Note that the contrast dependence inside the curly brackets clearly has a maximum, which explains the shape of the smooth
(red) curve in Fig 1.
   
Likewise,  to find $d^m$ and $V^m$ we can calculate the apparent magnitude of that fraction
$f$ of the galaxy-light lying inside the outermost detectable
isophote.$f$ is obtained simply by integrating Eqn (3) only to  $\Theta_c$ ,
corresponding to $I_c$ , in which case Eqn (6): \be f(\dmu)=1-e^{-N}.(1+N) \ee
where $N=0.92\dmu$ as always. So \be m=M+5logd(pc)-2.5logf(\dmu) \ee
And\be d(pc)=10\sqrt{f(\dmu})\times dex[0.2(m-M)] \ee
   
Thus the maximum distance $d^m$ to which the galaxy can be detected, without
exceeding the catalogue limit $m_c$ is \be
d^m(pc)=10.\sqrt{f(\dmu)}\times \left[\frac{1}{l_c}\right]^{0.5}\times
dex(0.2M) \ee where $l_c$ is the apparent luminosity corresponding to $m_c$ . In
its cubed form (18) yields $V^{m}$  the dashed (green) line in Figs 1 and 3
which reflects the monotonically falling nature of $f$ [ see (15)] as the
contrast $\dmu$, and hence $N=0.92\dmu$ vanish.
   
Having established the general shape of the red (smooth) and green
(dashed ) lines $V^\theta$ and $V^m$ in Fig 1 (and 3) what about their intersection
point P which will vary with their relative heights ? Dividing (14) by
(18) : \be \frac{d^\theta}{d^m}=\frac{Nexp(-0.5N)}{\sqrt{1-(1+N)exp(-N)}}\times \sqrt \frac{2}{\pi} \cdot \sqrt\frac{l_c}{I_c \theta_c^2} \ee where recall that $N=0.92\dmu$ . Fig 1(and 3) is  a plot
( of $(d^\theta)^{3}$ and $(d^{m})^{3}$ as a function of the SB contrast $\dmu$ i.e. of N.  

 It is evident from the above equation that the relative heights of the two visibilities can only be adjusted through the pure
number : \be\Gamma_c\equiv \frac{l_c}{I_c\theta_c^2}=\frac{-0.4dex(m_c -\mu_c)}{\theta_c^2}\ee determined by the catalogue parameters $m_c $ $\mu_c $ , and $\theta_c$ .  
Now it turns out (next section) that $m_c$ and $\mu_c$ are closely linked to one another by
photon statistics while $\theta_c$ is generally set by the telescope
resolution. Thus in practice $\Gamma_c$ has a narrow range. Hence the relative
heights of our smooth (red) and dashed(green) lines, which define the
Visibility Window, cannot sensibly vary by much, and in particular its
narrow aperture in contrast ( $< \pm$1.5 magnitudes) and its Wigwam shape
are more or less unavoidable as we shall see in the next section.

The net result of all the algebra is Fig 3 which looks very like the schematic Fig 1 but now is anchored in numbers, in particular the very narrow FWHM (2.5 mag), and the position of the Visibility peak P 3.5 mag above contrast zero. The actual curves and consequent Wigwam-shaped Visibility window were calculated from (14) and (18) assuming a value for $\Gamma_c$ ( eqn 20) of $\pi$ Ð typical of virtually all CCD surveys both in Space and from the ground (Fig 4, Sect 4). In looking at Fig 3 it is worth anticipating two points: (i) The contrast-zero point is locked to the absolute sky brightness being, for distant galaxies, about 5 mag dimmer than the sky in Space, 6 mag dimmer than the (brighter) sky on the ground; (ii) The 3.5 mag contrast at the peak refers only to the central brightest point of an Exponential galaxy. Most of it huddles much closer to the sky (Fig 9). [A more detailed
description of Visibility Theory can be found in Disney and
Phillipps (1983), though it lacks the vital arguments of the next
section; also see Ellis G.F.R et al. 1984)]
    
\begin{figure*}
\begin{center}
\includegraphics[width=3.5in]{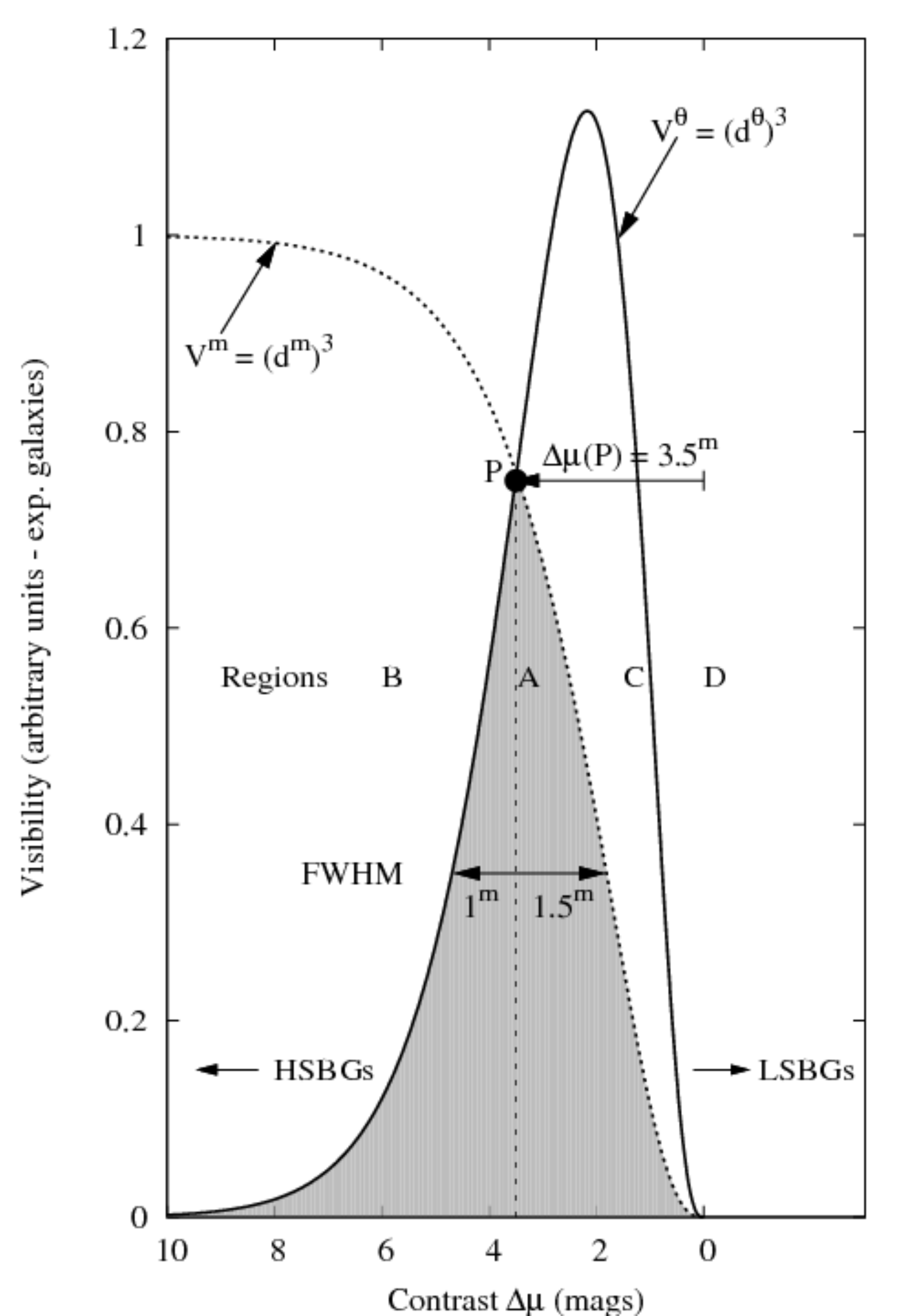}
\caption{The calculated Visibility Window for Exponential
  galaxies. The vertical scale shows the relative volumes within which
  galaxies with different surface-brightness-contrasts to the
  background (plotted horizontally) can be detected. Following the
  usual convention this contrast $\dmu$, in mag, is plotted from
  right to left with high surface brightness, i.e. high contrast
  galaxies to the left, low surface brightness galaxies to the
  right. The maximum heights of the two curves $V^m$ (dashed or green)
  and $V^{\theta}$ (smooth or red) assume a sample for which
  $\Gamma_c=\pi$ , typical of all Exponential galaxies, save those
  hundreds of pixels across. This is a typical Wigwam Diagram for the
  Visibility of galaxies of all kinds ( see later). Since the
  vertical scale is arbitrary the Wigwam Diagram is valid irrespective
  of Absolute Luminosity, just as it is valid irrespective of the
  absolute survey depth ( deepest isophotal level $\mu_c$ ) because
  the horizontal axis is given only in contrast $\dmu \equiv (\mu_c -
  \mu_0)$ where the latter is the central SB, measured in
  magnitudes. To be detected galaxies must lie inside the Wigwam, the
  shaded area marked A, which we call ÔThe Visibility
  WindowÕ. Note how narrow it is, with a FWHM of ~ 2.5 magnitudes
  with a peak P at a contrast of 3.5 magnitudes .Because the Window is
  so narrow, redshift dimming will quickly move galaxies rightward and out of sight
  into regions C and even D. For future reference  note
  that even the $V^\theta$(smooth or red) curve, by itself, has a FWHM
  of only 3 magnitudes.}
\label{fig3}
\end{center}
\end{figure*}
   
\section{IMPRISONED BY LIGHT}
    
The precise shape and location of the Visibility Window for a given
galaxy survey will depend on the relative heights of the red (smooth)
and green (dashed) lines in Fig 3 which in turn depend on the number $n_p$
of photons gathered by the detector per square arc sec. For instance
if the smooth(red) line $V^\theta$ is lower than the green(dashed) $V^m$ at all
contrasts $\dmu$, then it is only the smooth(red)$V^\theta$ , with its FWHM and peak,
which define the Visibility Window Ð which then might be quite
different from the Wigwam calculated in Fig 3. It turned out that the
relative heights were determined by the pure number $\Gamma_c$ , but what
determines $\Gamma_c$? This is the important step in the argument missing from
the 1983 paper.
    
Imagine a roughly circular source $\Theta$ arc sec in diameter where the
detector has collected  $n_p$ /photons sq arc sec. The signal from the source

\be l = \frac{\overline{I}}{I_{sky}} \times  \pi ( \frac{\Theta}{2})^2 \cdot n_p \ee

 where $\overline{I}$ = a level of signal from the source in photons collected/ sq arc. sec. averaged over the whole source-area. $I_{sky}$ is the foreground sky level and  $n_p$  is the total accumulated  signal in photons/ sq arc sec. Then from ( 15)
\be \Gamma_c \equiv \frac{l_c}{I_c\theta_c ^2}=\frac{\overline{I}(\pi/4)\cdot\Theta^2}{I_c \theta_c^ 2}\ee
       
where $ I_c $ is the average level of signal within the outermost detectable contour in photons/ arc sec sq.Thus for the limiting case of the
smallest  sources detected in the catalogue $\Theta\rightarrow\theta_c$ 
And\be\Gamma_c\rightarrow\frac{\pi}{4} \cdot \frac{\overline{I}}{I_c}\ee
        
      What are $\overline{I}$ and $I_c$ ? They will be set by signal-to-noise considerations. For the whole source the signal is given by (21) while photon-noise  from the sky(assumed brighter than the source) is
$\sqrt {\pi n_p (\Theta/2)^2}$
       
so the S/N of the whole source $\equiv  \sigma_m =(\overline{I}/I_{sky})\cdot \sqrt{\pi n_p (\Theta/2)^2}$ or  
\be\frac{\overline{I}}{I_{sky}} =\frac {\sigma_m}{\sqrt{n_p \pi (\Theta/2)^2} }\ee
       
       $I_c$ will likewise be set by the S/N in the outermost isophote
which we will assume  is $q\Theta$ wide (defining q). The signal in that outer isophote $=I_c \cdot \pi \Theta\cdot q\Theta$
       
The noise in it $=\sqrt{ \pi\Theta q \Theta n_p}$
       
And so  $ S/N\equiv \sigma _\theta =(I_c/I_{sky})\cdot\sqrt{\pi \Theta q\Theta n_p}$
       
Thus \be \frac{I_c}{I_{sky}}=\frac{\sigma_\theta}{\sqrt{\pi \Theta q\Theta n_p }}\ee
       
We can thus insert (24) and (25) into (23):
\be \Gamma_c =\frac{\pi}{2}\cdot \sqrt {q}\cdot \frac{\sigma_m}{\sigma_\theta}\ee
       
where q, defined to to be the width of the outer isophote as a fraction of the diameter, depends only on the sizes of the sources at the limit of detection.
       
From (26) it becomes clear that estimating $\Gamma_c$ , and hence the location of
the Visibility Window, relies on picking appropriate values for the
two limiting signal-to-noise ratios, $\sigma_m$ referring to a whole image, and $\sigma_\theta$
to its outer isophote, which define the catalogue. If (but see later)
the noise is dominated by photon-statistics, i.e. is binomial in
nature, there is a rational way to select those ratios. They must be
just high enough to avoid a significant number of Ôfalse
positivesÕ. As is well known in the Binomial  situation the probability of a single false positive
(i.e. single-tail) is given in Table 1.
\begin{table*}
 \centering
 \begin{minipage}{140mm}
  \caption{Probabilities of a false positive (Single tail)}
  \begin{tabular}{@{}lllllll@{}}
  \hline
   $\sigma$    & 1 & 2 & 3 & 4 & 5 &6 
   \\            
   \hline
   Prob & 0.16 & .03 & .0015 & $3\times 10^{-5}$ & $3\times 10^{-7}$ & $10^{-9}$
   \\
   \hline
\end{tabular}
\end{minipage}
\end{table*}

Thus in a survey of a $ \emph{single} $ CCD frame ($\sim 10^7$ pixels) a formal choice of a discriminating $ \sigma_m$ = 5 should eliminate all but a handful of false positives. For a survey consisting a fair number of CCD frames a S/N
of 6 to 7 would be safer. 

 The case for $\sigma_\theta$  is different. The source has
been selected; one needs only to be reasonably certain that the
apparent outer isophote is real, i.e. that itÕs probability as a
false-positive is less than say 5 or 10 percent, in which case $\sigma_\theta \sim 1.5$ should suffice. 

 All that remains uncertain in (26) is q. Now for
small extended sources containing only 10 to 20 optical pixels $\emph{altogether}$,
i.e. galaxies at the limit of detectability as such in Hubble Deep
Fields, q $\leq$ 1/3 ,i.e. the diameter must be $\geq$ 3 times the outermost isophote
-width.  

Thus for HDFs $\Gamma_c \approx \frac {\pi}{2} \cdot \sqrt {3^{-1}} \cdot \frac {7}{1.5} \approx 4.2$. However ,if we had the luxury of a
catalogue comprised of large galaxies $ > 100 $ pixels across q$\rightarrow 10^{-2}$ and hence $\Gamma_c\rightarrow0.5$.

We can summarise the bounds on $\Gamma_c$ as follows:

 \be  0.5 < \Gamma_c < 5 \ee

where very large galaxies having hundreds of resolution elements per diameter are on the left , and extremely small galaxies  having 3 to 5  are on the right. 

$\Gamma_c$ is so important because it
determines the crossover point P, i.e. $\dmu(P)$   and thus the nature of the
Visibility Windows in diagrams such as Fig 1 and Fig 3. One can find $\dmu(P)$ 
for a given $\Gamma_c$ by equating $V^\theta$ to $V^m$ and solving Eqn (19) for $\dmu$ .  There are no
solutions for N $< 2$ (i.e.  $\dmu < 2/0.92 = 2.2$) because then the dashed
(green) line always passes above the smooth (red) line , and none for N $> $4
( $\dmu > 4.5$) because $\Gamma_c$ must be $< 5 $ [see Eqn. (27)]. In between there is a rather smooth, almost linear transition which passes through the $(\Gamma_c, \dmu(P))$
points (2.5, 3.0), (3.2, 3.5) and (4.9, 4.4). See Fig 4. Thus Fig 3 ($\pi$ , 3.5) is
completely typical of all but the largest galaxies with hundreds of
detector pixels per diameter.
           
\begin{figure*}
\begin{center}
\includegraphics[width=4in]{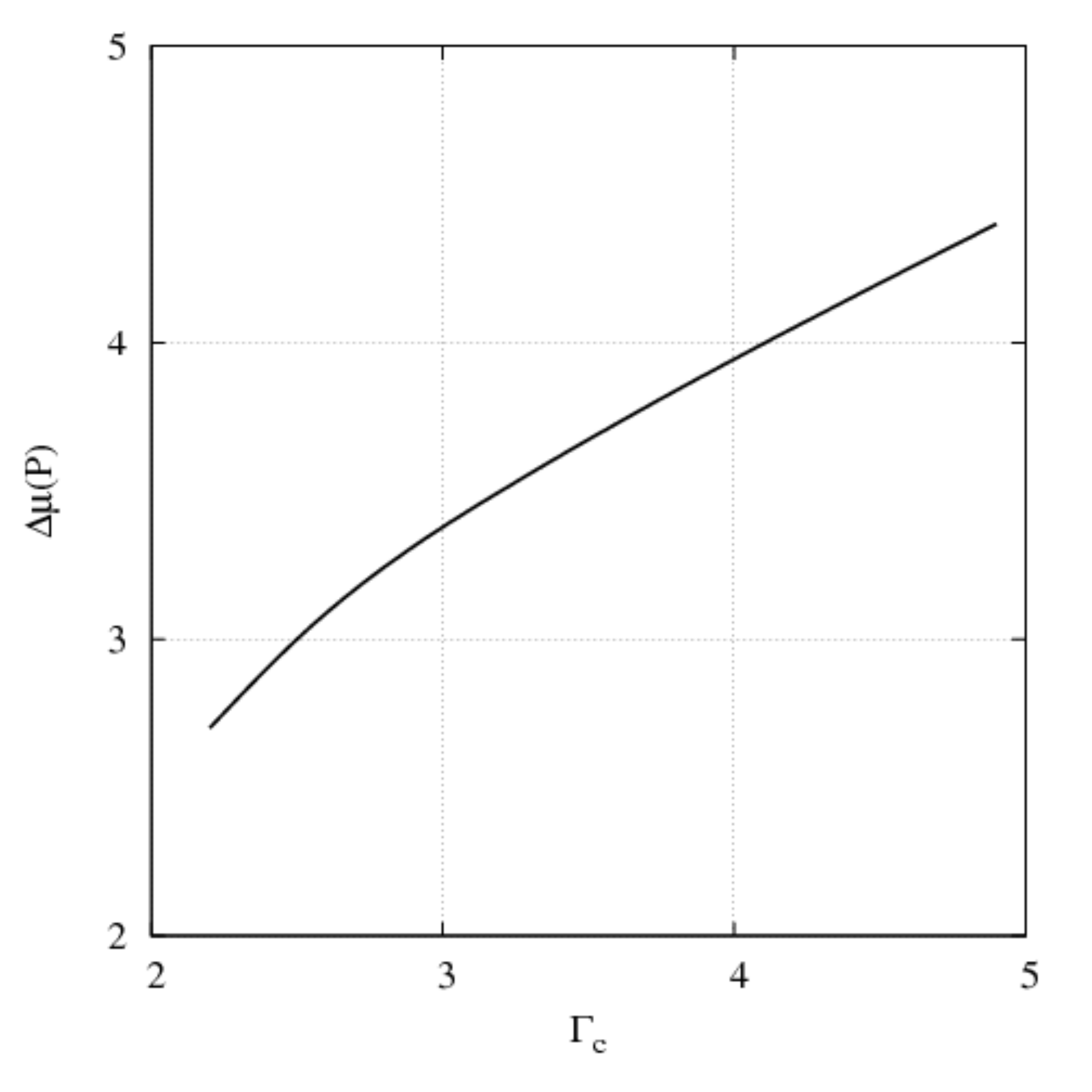}
\caption{ The surface brightness contrast $\dmu(P)$, in mag, at the
  peak of the Visibility Wigwam (see figs 1 and 3) for Exponential
  galaxies in surveys with different values of the pure number
  $\Gamma_c\equiv l_c/I_c\theta_c^2$ (see Eqn 20). Values calculated
  numerically using the procedure discussed below Eqn. 27. For the
  smooth (red) and dashed (green) curves to cross $\dmu(P)$ must be
  $>$ 2.2 (see Eqn. 27) and $\Gamma_c$ must be $<$ 5 for the smallest
  galaxies. Thus the practical range in $\dmu(P)$ for nearly all
  surveys is narrow (3 to 4 mag) .  Thus the Visibility curve shown in
  Fig.3 with $\Gamma_c=\pi$ and $\dmu(P)=3.5$ is very typical for
  Exponentials. Only very large galaxies with hundreds of
  pixels/diameter have $\Gamma_c$ Õs less than 2.2. For them the
  smooth (red) $V^\theta$ Visibility curve is all that applies. [NB
    The algebra for Ellipticals is slightly different but otherwise an
    identical procedure leads to a crossover at $\dmu(P)$ = 10.4 mags
    for a $\Gamma_c$ of 5 (See Figs 7 \& 9)] }
\label{fig4}
\end{center}
\end{figure*}   
            
We can summarize a situation, which is very much simpler than it might have
been, as follows. A search for Exponential galaxies with a CCD
detector will have a Visibility diagram much like Fig 3, i.e. a Wigwam
Diagram. Such galaxies will only be found in a narrow Visibility
Window centred at a contrast of between 3 and 4.5 magnitudes, and the
FWHM of that window will be 1 mag to the high SB side, 1.5 mag to the
low SB side, making a total FWHM of only 2.5 magnitudes in all. For
very large ($ >$ 100 detector pixels in diameter) galaxies the situation
is qualitatively different. Only the (smooth, red) curve in Fig 3 then
matters, in which case the Visibility Window will be centred at a
contrast of 2.2 magnitudes with a FWHM of 3 mags, 2 on the high SB
side, 1 on the low ( as assumed by Disney 1976, and reviewed by Impey and Bothun 1997).
\newline

To emphasise how implacably we are imprisoned in our local cell of
light let us try to calculate a way out of it.  

The number of galaxies $\hat{N}$
of co-moving density $\phi$ we will detect in a survey covering solid angle $\Omega$
will be

       \be\hat{N}\propto \phi \frac{\Omega}{3}\cdot d_{max}^3\ee

To move rightwards in Fig 3, i.e. to lower SB, it is the green
(Dashed) line $V^m$ which matters, i.e. the galaxy's apparent magnitude
must exceed the level of sky-noise by some discriminating S/N factor $\sigma$:

 For a circular source of diameter $\Theta$ arc sec in an observation containing $n_P$ photons per arc sec$^{-2}$ (24):
 
  S/N $\equiv \sigma = [\overline{I} /I_s] \cdot \sqrt { \pi/4\cdot \Theta^2 n_p}$
 
 Now $\Theta\propto $ R/d  (R = radius, d = distance) 
 Thus  \be \Theta \propto \frac{\sqrt{L/ \overline{I}}}{d} \ee
 
 or \be d_{max} = \frac {1}{\sigma} \cdot \frac {1}{I_{s} }\cdot \sqrt{L\overline{I}}\cdot\sqrt{n_P}\ee
 
 Now \be \Omega= \frac {TW}{t}\ee
 
 where T is the total survey time, t the dwell time per frame, and W the solid-angular area of the field of view of a single frame.This last equation is obvious but crucial because it argues that increasing dwell-time t in order to search for lower SB galaxies will not be so productive because it will, at the same time, reduce $\Omega $ and hence the Volume that can be explored.
 
 Thus \be \hat{N} = \frac {1}{3} \left(\frac{TW}{t}\right) \cdot  \frac {1}{\sigma^3 I_s ^3}\cdot (L\overline{I} )^{3/2} \cdot n_P ^{3/2}\ee
 
 But \be n_P = I_s [D^2tQ\Delta\lambda]\ee
 
 where D = telescope diameter, t= dwell time, Q is quantum efficiency of the system and $\Delta\lambda$ is the bandwidth of the detector in Angstroms  say, and we have assumed, for low SB galaxies, that most of the collected photons come from the sky.
 
 And so putting all together:

 \be
            \hat{N} \propto \phi \cdot T\cdot \left(L \frac {\overline{I}}{I_s}\right)^{3/2}
           \times [D^{3} t^{0.5} ]\cdot  \{WQ^{3/2}\} \ee

a very important relation which neatly separates the galaxy
properties( ), the survey properties[ ] and the detector power $\{ \}$.

 From (34) we infer:

(a) To acquire a certain number galaxies of SB $\overline{I}$:
            
          \be t\propto \left(\frac{I_s}{\overline{I}}\right)^3\ee
            
i.e. for a drop in SB of 1 mag the dwell time must be increased by 3
mags, or a factor of 16. Thus to escape entirely out of our Visibility
Window (FWHM~ 2.5 mags) on the low SB side, we would need to increase
dwell times by 2.5$\times$ 3 mag or a factor of a thousand! $\emph{We truly are
imprisoned in our lighted cell}$. [Indeed the situation may be even
  worse than we have supposed. Thus far we have assumed that the two
  galaxy parameters L and $ \overline{I}$ are independent Ð which may not be
  true. In so far as we can disentangle the two Ð which requires a
  sample selected by non-optical ( e.g. 21-cm.) means, the suggestion
  is that $\overline{I}\propto (L)^{1/3} $(Garcia-Appadoo et al 2009, Chang et al 2011). If that is true then to see dim
  objects the dwell time t must increase not as $(\overline{I}/I_s)^3$but as$(\overline{I}/I_s)^6$ ! See also Sect
  IX.]
            
(b) The detector figure-of-merit $\{ \}$is higher for CCDs than for Schmidt
photographic plates $\{36 sq. deg, Q \sim 0.01\}$ provided the CCD (Q$\sim$0.5)
has $>$ 2000 pixels a side. The grasp of any survey, by (34):
\be\propto T\times \left ( D^3
\cdot t^{0.5}\right)\times WQ^{3/2}\ee

which means that 1-month-long CCD surveys with 4-metre class
telescopes will be an order of magnitude less effective for finding
LSBGs than the combined Schmidt surveys covering the whole sky. But if
photon-counting were the whole story then the Sloan DSS ought to beat
the Schmidts by a factor of between 5 and 10, despite its very short
dwell time $\sim$ 100 secs. Unfortunately very low SB galaxies can only be
detected if they look apparently large [Eqn. 24] when the unevenness
of the sky-background, not its photon-statistics, becomes the
predominant source of noise (Sabatini et al. (2003).
            
(c) The one ray of hope is the $D^3$ in Eqn. (36). Alas large telescopes
produce larger images which over-fill the CCD-detector pixels for
diameters $>$ 2 to 3 metres on the ground because optics cannot be made
arbitrarily ÔfastÕ. In that case $W\sim D^{-2}$ requiring $D \sim \overline{I}^{-3/2}$ for a given 
 $\hat{N}$
. Telescope diameters of 100 meters would be needed (see sect 6) to move one
window-width dimmer than we can see now. Only low-noise, high-quantum-efficiency detectors of far
larger physical size than CCDs offer any prospect of escape. 

(d) Thus far we have estimated everything in terms of $\dmu\equiv\mu_c -\mu_0 $ where $\mu_c $ is a so far numerically unspecified SB , presumably connected to the sky-brightness $\mu_{sky} $ by signal-to-noise considerations. Recall that it is the SB of the outermost detectable contour in a galaxy where that contour width is, by definition, a fraction q of the galaxy's total angular diameter $ \Theta$ . We found ( (25)) that

\be \frac {I_c}{I_{sky}} = \frac {\sigma_\theta}{\sqrt{\pi  q}\cdot \sqrt{\Theta^2 n_P}} \ee

For the smallest galaxies detectable in a survey $\pi q \sim 1$ so:

\be\frac{I_c}{I_{sky}} \sim \frac{\sigma_\theta}{\sqrt{N'}} \sim \frac {1.5}{\sqrt{N'}} \ee

where $N' \approx \Theta^2 n_P $ is the total number of photons collected, largely from the sky, from  $ \emph{ an\  area \ equivalent\  to\  the\  area\  of \ the \ whole\  source}.$

So far as Space is concerned the high resolution of HST means that $\Theta $ is very small for faint galaxies$ ( \leq 10^{-1} arc sec)$ so that extremely long integrations (tens of orbits) are needed to achieve N' s as high as $10^4$ photons. It follows from (38) that

\be\mu_c\approx\mu_{sky} + (4.75 \ mag) \ee

Thus in Space $\mu_c$ is locked to the sky brightness. The position of the Visibility Window up there is not merely defined in terms of contrast  but is in practice locked in $\emph{absolute}$ surface-brightness terms too.

     The same is true on the ground though the argument is slightly more subtle. According to (34), for a given $\hat{N}$ :
     
     \be\frac{\overline{I}}{I_{sky}}\propto \frac {1}{L}\cdot\frac{1}{t^{1/3}}\cdot\frac{1}{D^2}\cdot\frac{1}{T^{2/3}}\cdot\frac{1}{W^{2/3}}\cdot\frac{1}{\phi^{2/3}}\ee
     
     from which it might seem that a sufficiently long dwell-time t might lead to the detection of arbitrarily dim galalxies. Not so because (40) ignores Tolman dimming. It is easy to show that such dimming modifies (40) to
     
     \begin{eqnarray} \frac{\overline{I}}{I_{sky}}\propto  (1+z)^2 \times \frac {1}{L}\cdot\frac{1}{t^{1/3}}\cdot\frac{1}{D^2}\cdot\frac{1}{T^{2/3}}\cdot\frac{1}{W^{2/3}}\cdot\frac{1}{\phi^{2/3}}\nonumber \end{eqnarray}
     
     If, because of the $(1+z)^2$ term, you cannot afford z to rise above 0.2 say then to find sufficient $(\hat{N}) $ galaxies you must increase the area coverage $\Omega$ in (28) by taking a number of frames $ \Omega/W=T/t$ . Putting in reasonable values for W (one CCD) and $\phi$ it then transpires that to find a handfull of low SB $ L_*$ galaxies within $ z \approx 0.2 $ would require:
     
     \be\frac{t}{T}\leq 10^{-4}\ee
     
     Now a pixel-matching (i.e. 2 to 3 M) telescope collects $n_P\sim10$  sky-photons sec$^{-1}$ arc sec$^{-2}$,  so in an area of a 3 by 3 arc sec galaxy  $n_P\sim 10^2$  sky photons/sec so that in a long campaign lasting $T=10^7$~sec  (i.e  $ t \ \approx 10{^3} sec) $ $N'= 10^5$ photons/galaxy-area. So from (38)
     \be \mu_c\approx \mu_{sky} + 5.8 \ mag\ee
     
     which again is locked in absolute terms to the SB of the terrestrial sky (which at most wavelengths is $\emph{at least} $ one mag brighter than it is at HST).
     
     The fundamental point is that the Visibility Wigwam diagrams are fixed not only in contrast terms but $\emph{in absolute  surface-brightness terms as well.}$

An alternative way to look at the matter is to investigate how the dimmest galaxy (SB $\sim I_{min}$) one can detect improves with telescope diameter D. On the ground, because of the pixel-matching problem, $I_{min}\propto \left(DA\right)^{-2/3}$ where A is the physical area of the detector. In Space pixel-matching isn't an issue because the diffraction-limited angular resolution $ \delta\theta \propto D^{-1} $ . But then, for a fixed number P of pixels , the survey area $\Omega\propto P\left(\delta\theta\right)^{2}\propto PD^{-2}$ and so again $I_{min}\propto D^{-2/3}P^{-1} $. In other words the telescope costs of escaping from the Visibility Window, be it in Space or on the ground, become exorbitant. A factor 10 improvement in $I_{min}$ would imply an increase in telescope diameter of $10^{3/2}$ and hence in costs C of $10^{3\gamma/2}$ if $C\propto D^\gamma$. Since $\gamma$ is usually reckoned to lie between 2.5 and 3, and certainly above 2, vast sums would be needed.

For all practical purposes then we $\emph{are}$ implacably imprisoned in our cell of light. Classes of low SB galaxies unresolved into stars, which cannot already be seen in Schmidt surveys, are beyond hope of discovery by optical means alone.  It follows that  large hidden populations of low surface brightness galaxies, both near and far, cannot be ruled out by optical observations alone. This is a much stronger statement than could have been made before and it relies on the  arguments which led to eqn.(26)

            \section{HOW GALAXIES SINK FROM SIGHT}
            
The Visibility Window depicted in fig 3 is immutable, mathematical and
pinned in local coordinates because it shows the contrast to oneÕs
local sky, be it on the ground or in space.             
What we need to calculate next are the properties, in particular the sizes and intrinsic SBs ,of the  kinds of galaxies, seen at
different redshifts, which will make it through that narrow window,
particularly near its peak, taking into account the Tolman effects described above, which both dim a galaxy and increase its apparent size.

.
            
The $(1+z)^{-4}$factor rapidly becomes very significant by comparison with the
narrow FWHM (2.5 mag) of the Visibility Window. Even at z = 0.5 many
of the most Visible galaxies that were in region A (Fig 1) at low
redshift would be translated into region C and be far too dim to
see. They have Sunk. Their SB contrast now becomes:

\be \dmu'\equiv \mu_c- \mu_0'=\mu_c-[\mu_0+10log(1+z)]=\dmu-10log(1+z)\ee

which implies that even galaxies at the peak of the Visibility Window
at low redshift [ where ~$\dmu\sim 3.5$] will have zero contrast  $\dmu'$ i.e. will cross the green (dashed) line [Fig 1] and vanish entirely by a redshift of
1.2 To delineate that green (dashed) line recall that the fraction of
light detected above the outermost isophote $\mu_c$  is given by Eqn 15 .Figure
5 depicts $f(\dmu)$ . More than 50 per cent of the light from a galaxy that
would be at the peak nearby, has already been lost at redshift 0.5, 82
per cent at redshift 1, and all by 1.2. These figures alone are enough
to query the feasibility of trying to study galaxy evolution by using
deep fields.
            
\begin{figure*}
\begin{center}
\includegraphics[width=4in]{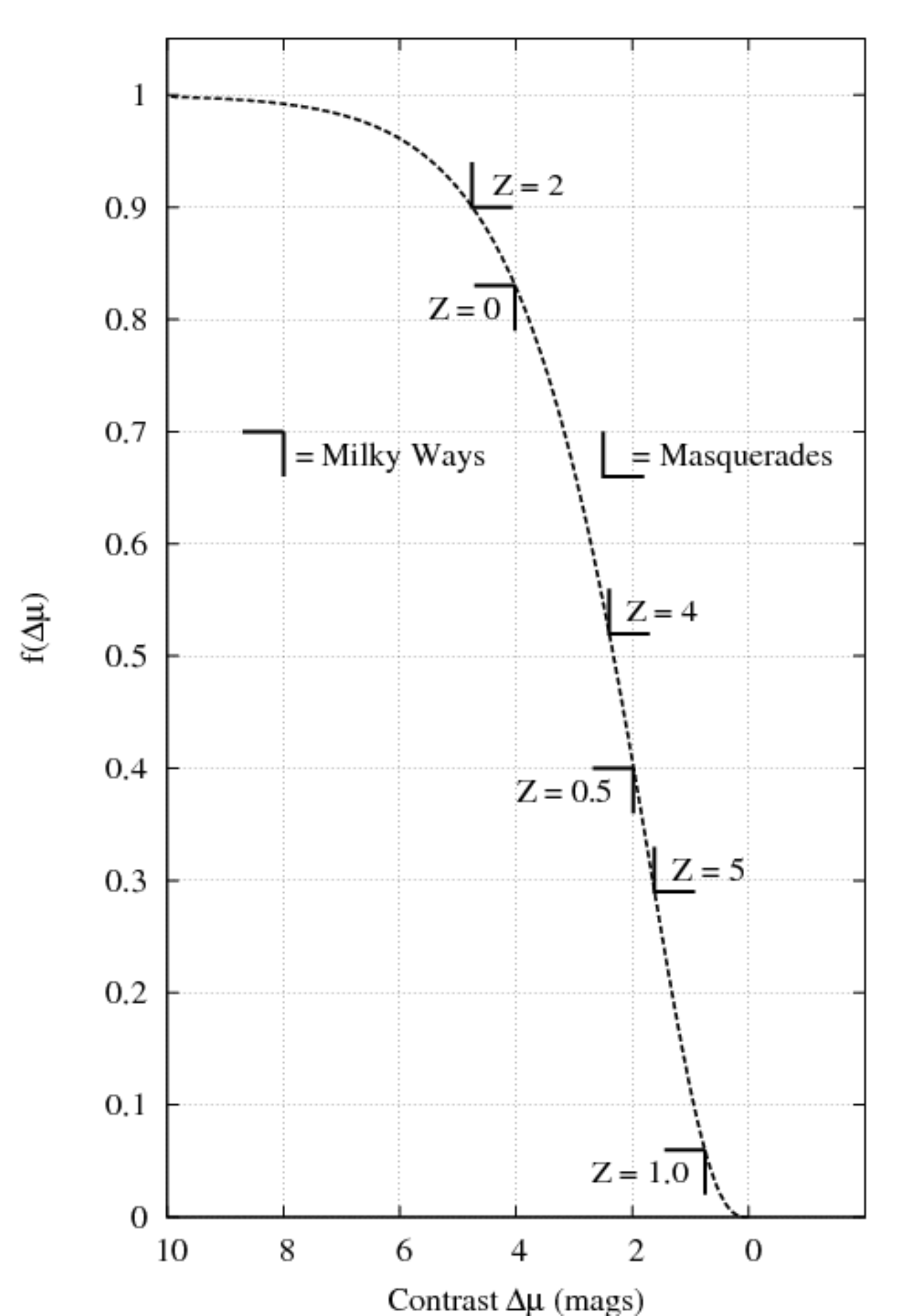}
\caption{The curve shows $f(\dmu)$ , the fraction of an Exponential
  galaxy's light seen above the outermost detectable isophote
  $\mu_c$, plotted against the galaxy's contrast $\dmu \equiv(\mu_c
  -\mu_0)$ in magnitudes. It is calculated from Eqn 15 with $\dmu$ modified using Eqn (43). Thus a galaxy like the MW with an optimal
  contrast $\dmu(P) \approx 3.5$ mag at z=0 has an $f(\dmu)$= 0.83 there, as
  shown by the tick mark. By the time it is removed to z=0.5 ,
  $f(\dmu)$ has dropped to 0.4, by z=1.0 to .06 and it disappears
  altogether at z=1.2 due to Tolman dimming. By contrast the tick
  marks to the RHS of the line show an $L^\star$ galaxy 9 mag higher in
  SB Ð a so called 'Masquerade'. Even at z= 2 ninety per cent
  of its light is still visible, and by z = 5 nearly 30 per cent is
  still left. It only sinks completely at a redshift of 7.}
\label{fig5}
\end{center}
\end{figure*}   
             
The galaxies that will appear instead at the peak of the Window will
be, as always, those with an apparent contrast $\dmu'$ of $\sim$ 3.5 mag. In other
words their intrinsic SBs will be given by [ see (1)]

\be\mu_0 \approx \mu_c -3.5-10log(1+z)\ee

or, at z = 0.5, 1.8 mag more brilliant than optimally Visible
galaxies nearby to us today at low redshift ($\mu_0\sim 21.5 V\mu$) and 3.0 mag more
brilliant at z = 1. Indeed if one examines the Visibility Window
(Fig. 3) one sees, down at the FWHM, that the z =1 galaxies now in the
window must have emerged, or ÔsurfacedÕ from Region B where they
would be practically invisible at redshift zero.

How can redshifting, and hence dimming a galaxy render it more
Visible? What the Visibility Window illustrates is the $\emph{relative}$
Visibilities of galaxies with different SBs. Rare but high-Visibility
galaxies can be seen at great distances, common but low-Visibility
galaxies may rarely turn up close enough to us to be noticeable in
surveys. If now we remove the Local population to redshift 1,
virtually all the previously prominent galaxies will sink below the
sky Ð thanks to Eqn. (43). Our high SB specimen therefore has much
less competition, and is correspondingly more prominent. In addition
it has gained through aberration. Whereas removing it to z = 1 would
normally render it too small to seen as a galaxy  (i.e. $\theta < \theta_c$ )  aberration may return it
from the invisible region B into the visible window A.
             
In qualitative terms then, removing any population of galaxies to
higher redshifts will drastically alter their relative Visibilities,
so that the previously prominent specimens sink partially, or wholly,
out of sight, to be replaced there at the peak of the window by
intrinsically more brilliant galaxies that were relatively
inconspicuous at low z because of their small apparent sizes. It is
time to make things quantitative.
             
Begin by calculating the apparent magnitude m(z) of galaxies that have
peak Visibility (i.e. $\dmu'\equiv\mu_c -\mu_0)\approx 3.5$
at redshift z taking into account both Tolman
dimming, and Cosmology.  Apparent luminosity:

\be l(z)=\frac{L}{4\pi d^2(z)}\cdot\frac{1}{(1+z)^2 }\cdot f(\dmu') \ee

where $f$ is the fraction of the light seen above the sky [Eqn.(15)] and $\dmu'$
has been adjusted for redshift according to (43). Convert to
magnitudes, with distances in Mpc. and

\be m(z)=5[log d(z)(Mpc)+6]-5+5log(1+z)-2.5log f\ee 
             
d(z) is the ÔproperÕ co-moving radial distance defined such that
the co-moving volume element out at z is $ \Delta\tau\equiv \frac{1}{3}d^2(z)\cdot\Delta\Omega\cdot \Delta d(z)$ corresponding to solid angle $\Delta\Omega$
.[We donÕt need to employ the concepts of ÔLuminosity distanceÕ
  or ÔAngular-size distanceÕ because we incorporate the (1+z)
  factors directly into equations such as (45) and (46).]
             
Cosmology now enters only through the functional dependence of the
co-moving distance d(z) on z. It can be a complicated function
depending, as it may, on the various model-parameters  $\Omega_M,\Omega_\Lambda ,\Omega_0, H_0 $ and so on . Here
we use the Ôempty-universeÕ approximation:

\be d(z)=\left(\frac{c}{H_0}\right) ln(1+z)\ee

 because it is simple , and closely approximates the currently
fashionable $\Lambda CDM$ model. Between $0.1 < z < 10 $ the discrepancy is a maximum of 12 per cent (at z=1) and for most of the range is much less [ as
  can easily be checked using Ned WrightÕs very useful on-line
 ÔCosmology Calculator (Wright 2006)]. Given uncertainties as to
which is the correct model, and K-corrections, dust and Evolution, this approximation is more than satisfactory. 
\\

Incorporating (47) into (46)

\begin{eqnarray} m(z)=M+25 - 2.5logf(\dmu')+5log\left( \frac{c}{H_0}\right) \nonumber\\ +5log[(1+z)\cdot ln(1+z)]\end{eqnarray}

    Likewise to find $\theta"(z)$  use (47) for d(Mpc) and (12) becomes

    \begin{eqnarray} \theta"(z)=\sqrt {\frac{2}{\pi}}\cdot \{ 0.92\dmu' exp (-0.2\dmu')\} \cdot(1+z)\nonumber\\ \times \frac{H_0/c}{ln(1+z)} \cdot dex(-5)\cdot dex[0.2\mu_c]\cdot dex[-0.2M] \end{eqnarray}
     
where the (1+z) term incorporates the aberration.
             
\begin{figure*}
\begin{center}
\includegraphics[width=4in]{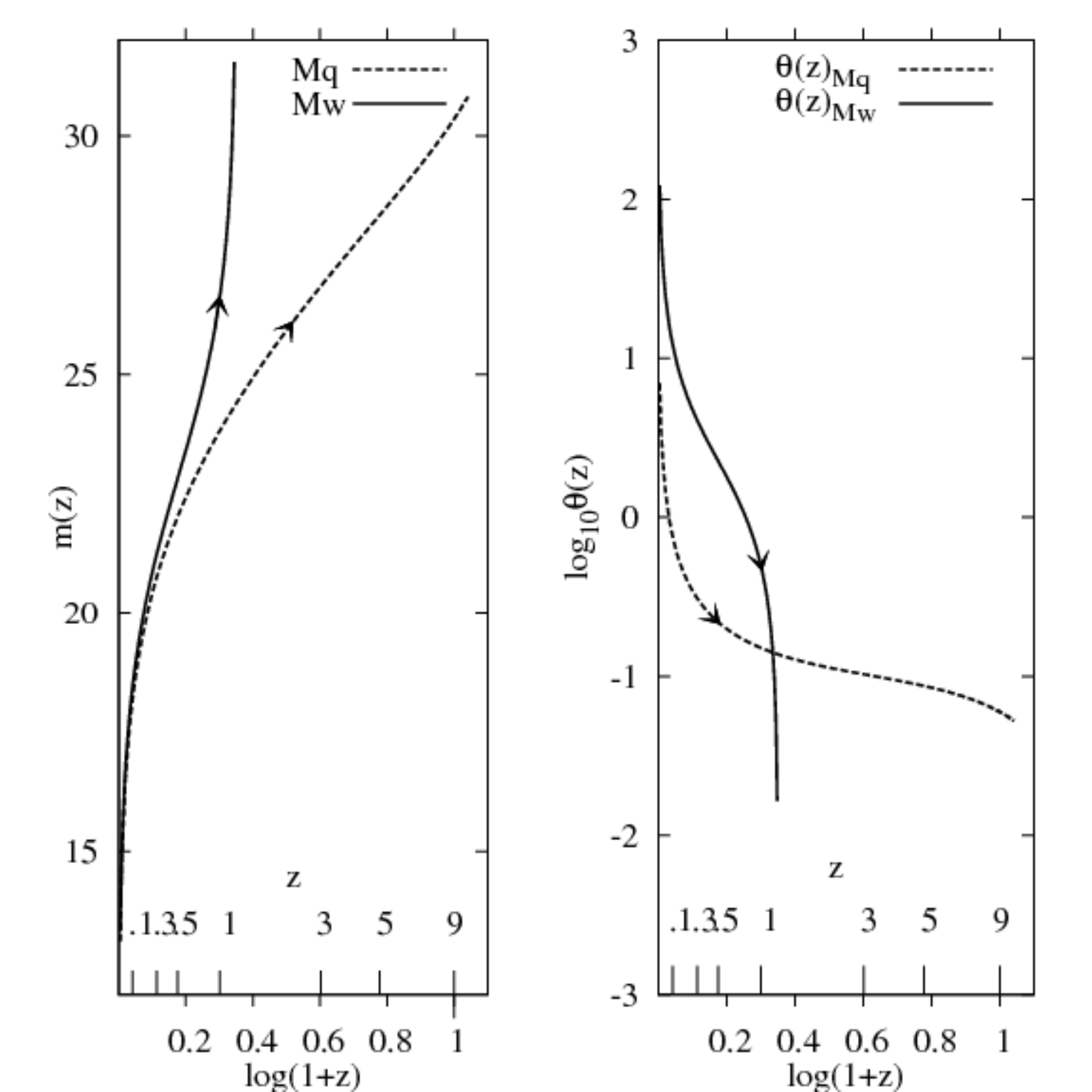}
\caption{The appearance of Exponential galaxies as a function of their
  redshift z. The apparent magnitudes m(z)(left panel) and angular
  sizes $\theta(z)$(diameter arc sec) (right panel) are shown for two
  objects of very different intrinsic surface brightness: a Milky-Way
  labelled Mw and a Masquerade labelled Mq which is a hypothetical
  galaxy of the same Luminosity but which is 9 mag (4,000 times)
  higher in SB, i.e. 9 mag more brilliant. The abscissa is redshift
  (plotted logarithmically, as are the other quantities).
Follow
  first the magnitude m(z) for the MW in the left panel. Because of
  redshift dimming, and shrinkage of its outer detectable isophotes
  against the sky, it rises more and more rapidly until, by z= 1.2 it
  will vanish even from the deepest Hubble Deep Fields [ magnitude
    limits $\approx$ 30 depending on colour ]. Follow second the
  angular size $\theta(z)$(right panel) of the Milky Way. It falls
  rapidly at first then slows as a number of factors including
  Aberration kick in, then falls again catastrophically as
  $z\rightarrow1$ when it loses contrast with the sky. It sinks
  completely out of sight when $z\rightarrow1.2$ [See Eqn. (43)].Now
  look at the Masquerade. Outer-isophote loss is negligible thus its
  m(z) ( left panel) increases more gradually with z so that its m(z)
  $\sim$ 29 at a redshift ~$\sim$ 7. It is still visible to HST out
  there. Its angular size $\theta(z)$ (right panel) which is barely an
  arc sec at low z, hardly changes with redshift, due to the (1+z)
  aberration term so it is distinguishable by HST as a galaxy even at
  z $\sim$ 7.}
\label{fig6}
\end{center}
\end{figure*}

Fig 6 employs the last two equations to investigate the appearances of
two galaxies at different redshifts. The first galaxy is a Milky Way,
the second a hypothetical galaxy of the same intrinsic luminosity but
with a SB no less than 9 magnitudes (4,000 times) higher.              
Notice first how quickly the MW sinks below the sky. By redshift half
56 per cent of its light has gone. By z = 0.9 the aberration cannot
compensate for the sinking of its outer isophotes , and by z=1.2 it
has sunk completely. One cannot expect to see healthy, i.e. more or
less complete MWs much beyond a redshift of 0.5.
             
Now look at the hypothetical Masquerade which would be only 330 pc in diameter. Being $(4,000)^{-1/2}$ smaller than the MW its
angular diameter at z $\sim$ 0.1 would only be 0.2 arc sec.
so unless it was  close ( $< $ 50 Mpc.) it would, from the ground, masquerade
as  a star, hence its name. However by a redshift of 4 aberration is kicking in,
while all the lower SB galaxies would have sunk, or be sinking out of
sight, so that by z $\sim$ 7 it would be the most Visible $L_*$ galaxy in sight
because its apparent SB would be $\sim$ 21.5 $\mu$, i.e. 3.5 mag brighter than
the SB limit $\mu_c$(See Sect VI). ItÕs angular size would be $\sim$ 0.6 arc sec
making it distinguishable to the HST as non-stellar, while its
magnitude would be $\sim$ 29.5 (Vega). And if z increased above 7 so would
its angular size , which would now be dominated by aberration. As
we shall see later it looks very like the z $\sim$ 7 galaxies being found
with the WFC-3 camera on HST.
             
We can summarise this section as follows. The sheer size of Tolman
dimming at the high redshifts accessible with HST makes it almost
certain that the population of galaxies we see out there is very
different from, and may not even be related to, our conspicuous
neighbours today. The narrowness of the Visibility Window (Fig 3)
compared to Tolman dimming is such that, without dramatic and
fortuitous amounts of Evolution (up to and beyond 9 mag), our
neighbours will fade dramatically beyond redshift 0.5 and sink
altogether below our local sky at z ~$\sim$1.2. Whatever the case nearby,
the distant (z $>$1) universe is almost certainly dominated by Sunken
galaxies that are invisible to us, sunken galaxies that would surely
alter our ideas on the star-formation history of the cosmos and its
re-ionisation, could we but detect them. Those who aim to decode these
matters by looking at the high redshift galaxies now visible with HST,
even to decode galaxy evolution beyond redshift one-half, must first
convince themselves that they are looking at our ancestors and not at
a very different, higher SB population, the one that is most visible
to us at that redshift, but which is inconspicuous nearby.

             \section{WHY HIGH REDSHIFT GALAXIES LOOK SMALL}
             
Technical developments, and in particular the fitting of the new WFC-3
camera to HST, make it almost trivial to find galaxies out to redshift
7, and perhaps higher. Its near IR sensitivity out to 1.7 microns, its
resolution there ( $\sim$ 0.1 arc sec.) and its field-of-view ( 4.8 sq arc
mins) conspire to make it $\sim$ 30 times faster for finding such objects
than previous space cameras like NICMOS. Such galaxies are observed in
their rest-frame UV (0.1 to 0.2 microns) where prominent breaks in
their spectra at Ly-$\alpha$ and at the Lyman limit make for fairly
unambiguous selection and photometric redshift measurements [
  e.g. Bouwens et al. (2010) ,Bunker et al (2010), McLure et
  al. (2010), Oesche et al ( 2010)]
             
If, as we are supposing, Surface Brightness selection through our
narrow Visibility Window dominates the appearance of galaxies out
there, one can make several strong predictions:
             
(a) All such high-z galaxies (indeed all Exponential galaxies in the
deep frames) should have a narrow range of $\emph {apparent}$ surface brightness
($\sim$ 3-4 mag).
             
(b) That range should be centred 3 to 4 magnitudes higher
than the limiting isophotal value for the observational data in
question.
             
(c) For the high-z galaxies Tolman dimming then implies
that their intrinsic SBs must be very high, $\sim$ 9 mags higher than
prominent galaxies nearby. This in turn implies that they must be
physically very small, otherwise they would be super-luminous.
              
(d) The apparent scale-length  for such exponential galaxies should appear to decrease with redshift in a well determined way, i.e.:
\be \alpha \propto (1+z)^{-1} \ee

(e) Either such super-compact galaxies have detectable descendants
nearby, or there must be some plausible mechanism for explaining their
absence ( next section).
              
Let us now compare these predictions with observations:

 (a) The
predicted constancy and scatter in SB is a direct consequence of the
previous three sections and hardly needs further discussion.
              
(b) Where do we expect the central peak of the distribution of the SB's of galaxies in a Hubble
Deep Field Window to lie?  At peak we know (Sect 3):$\dmu(P)=\mu_c -\mu_0=3.5$ .               
For small galaxies in Hubble Deep Fields Eqn (39):
\be\mu(P)=\mu_c-3.5\approx(\mu_{sky} +4.75)-3.5\approx \mu_{sky} +1.25\ee

Fig 2 shows the distribution of SBs in one of the Hubble Deep Fields. As can been seen it fits the prediction
of Visibility Theory very well because the sky brightness in the I band at 
Hubble is 22.5$\mu$ (Vega) .

 Given Tolman dimming, evolution and dust absorption,
all of which could be very large in these circumstances, especially in
the rest-frame UV, it is very hard to understand Fig 2 as other than
some kind of profound selection effect $\emph{operating in the observer's frame of reference}$, as the SPDH suggests it is.
                 
Oesch et al (2010) made a study of the structure of sixteen z$\sim$7
galaxies in this sample and report "With an average intrinsic size
0.7 $\pm$ 0.3 kpc these galaxies are found to be extremely compact, having
an observed surface brightness $\mu_J\approx 26 $ mag arc sec$^{-2}." $ Their fig 2 shows the half-light radii tracking Absolute
Magnitude so as to maintain that SB constancy. And in their fig 5 they
extend the sample to objects in the range z = 2 to 8, finding that the
measured (as opposed to corrected) UV SB Ð which they interpret as
a star-formation rate, Òremains relatively constant for the whole
redshift range from z $\sim$ 7 to z $\sim$ 4É. for galaxies with luminosities in
the range (0.3 to 1)$L_\star$.Ó
                 
(c) Size evolution. For galaxies to be seen in the Visibility Window
Eqn.( 44) demands that their SB
\be\mu_0\approx\mu_c -3.5-10log(1+z)\ee

                  Now the physical scale-length $\alpha \propto \sqrt{ L/I_0}\propto \sqrt{L/I_c} \cdot \sqrt{I_c/I_0}$
                  Thus for a given L and $I_c$

                  \be\alpha \propto  dex[-0.2\dmu] \propto [dex[-0.2(\mu_c -\mu_0)]\ee

                  therefore \be \alpha\propto dex[-0.2\{3.5+10log(1+z)\}]\ee 
                  
                  therefore\be\alpha\propto (1+z)^{-2}\ee

Thus the apparent scale-length will be, thanks to aberration, a factor
(1+z) larger, in which case we predict
                  
                  \be r_{1/2} \propto (1+z)^{-1}
\ee

Oesch et al (2010) compare their measured $r_{1/2}$s  with $(1+z)^{-m}$ over the range z = 2 to 8
and report m= 1.2$\pm$ 0.17 for luminous (0.3 to 1) $L_{\star, z=3}$ and m = 1.32 $\pm$0.52 for less
luminous (0.12 to 0.3) $L_{\star, z=3}$ galaxies respectively. "This is in agreement with previous
estimates where the sizes were found to scale roughly according to $(1+z)^{-1}$ "
( Bouwens et. al. 2004, 2006].
                  
Earlier Buitrago et al , 2008 ) measured 80 giant galaxies $(M> 10^{11}M_{solar})$ in the range 1.7$ <  $z $< 3$ using NICMOS and split the sample into Discs and
Spheroids. Discs are $2.6\pm 0.3 $ smaller than today, and Spheroids $ 4.3\pm0.7$ smaller. The
implied stellar densities in the past at  $\approx 2\times 10^{10} M_{solar} kpc^{-3} $  "are very high and as
high as Globular Clusters today." The Disc measurements too are
obviously consistent with $r_{1/2} \approx (1+z)^{-1}$ .

The same fall-off in physical size with
redshift proceeds all the way from z = 0 to to z = 7 with $R_{1/2}\approx(1+z)^{-1}$. For
instance Ryan et al (2011) have recently used WFC-3 in a 15-colour
search to isolate a sample of early-type galaxies, this time in the
interval z=1.6$\pm$0.6  , and compared their sizes with a very large sample of
equivalent SDSS galaxies at z $\sim$ 0.2. Again they parameterise the size
decline as $R_{eff} \propto (1+z)^{-m}$

                  and find m is mass-dependent this time. And

                  \be m(M_\star )=-1.8+log\left(  \frac{M_\star}{10^9 M_{sun}}\right) \ee

yielding m $\sim$ 1 for massive galaxies, and a statistical decline over
all objects of a factor 4 between redshifts 0 and 1.6
 
The decline in galaxy-size with redshift is the most remarkable and
consistent result in all the Hubble deep-field observations. It is
predicted, indeed demanded by the SPD hypothesis in which Tolman
dimming brings successively more compact galaxies to light at higher
redshifts, while sinking entirely out of sight their less compact
companions.
[NB :As an aside, the confirmation of the prediction that angular size should $\propto (1+z)^{-1} $ $\emph {might} $ be taken, a la Tolman, as rather direct evidence, so far lacking, that the Universe $\emph{is} $ expanding. It would rely on the assumption that intrinsic SB does not change much with redshift, as suggested by the archaeology of nearby galaxies.] 
                  
                  \section{WHERE HAVE THE DESCENDANTS GONE?}
                  
What happened to the spectacularly high SB galaxies we see back at
redshift 7? Have they evolved away either by mergers or passive
dimming, or are their descendants lurking around us today? We shall
argue that their direct descendants could well be present in our
neighbourhood but would have passed unnoticed because they would be
extremely inconspicuous, and for three different reasons. First their
compact physical sizes translate into angular sizes so small that
their Angular-Size ÐVisibility  $V^\theta $will be down on ÔnormalÕ
galaxies by a factor of 60 cubed. Secondly the dust-grains in such
compact objects would be on average 60 times closer to neighbouring
stars than they would be in a Milky Way galaxy today, and therefore be 
3,600 times more effective as absorbers. Very little of their optical
light would therefore escape making even the nearest of them
exceedingly faint. And finally, in co-moving terms, they appear to be
pretty rare which implies that the nearest of them would be far enough
away to make them, in terms of angular size, barely distinguishable
from stars. Take these arguments one by one:
                  
                  (a) Visibility:
                  
For compact objects it is Angular-Size Visibility $V^\theta$  which
counts. According to Eqn.(25) the most visible objects at redshift 7
must have a SB of 10 log(1+7) = 9 mag higher, and therefore a diameter
4.5 mag , or 60 times smaller than the galaxies in our vicinity. Thus
for a given luminosity their angular sizes would be 60 times smaller,
and their Visibilities $60^{-3}$ ~$\sim$ $10^{-5}$ less. They will be extremely inconspicuous.
                  
(b) Internal absorption. Large disc galaxies typically lose half their
light to internal dust absorption (Disney, Davies and Phillipps 1989,
Soifer Helou and Werner 2008), but compact galaxies ought to lose
vastly more. One will see into a disc-galaxy $\sim$ one mean-free-path $\lambda$
where $\lambda=1/n\sigma $ where n is the particle density, and $\sigma$ the particle cross-section
for absorption. Shrinking the disc radially by a factor 60 will
increase n by $\sim 60^2$, so the physical depth from which one could detect
light would, crudely speaking, decrease by the same factor, leading to
a loss of apparent luminosity $\sim 60^2 \sim  9$ magnitudes. In other words once a
disc becomes optically thick, compacting it further cannot increase
the apparent SB, and its apparent optical luminosity will decrease
with its area.
                  
(c) Rarity and apparent angular size. Mclure et al (2010) fit a
Schecter luminosity function to the faint end of the high z sample [where the statistics
  are "better"] and arrive at a co-moving density $\phi^\star =7\times 10^{-4} Mpc^{-3} mag^{-1}$ which is more
than an order of magnitude below the local value. Ignoring clustering
the expected distance to the nearest one from us ought to be $\sim$ $(3/4\pi \phi^*)^{1/3}\sim 7$  Mpc. [distance modulus $\sim$ 29] and as the physical size $\sim 20$ kpc / 60
~$\sim$ 300 pc., the angular size of the nearest one would be ~$\sim$10 arc sec,
whilst most would look stellar. They would also be very faint, even
the nearest z ~$\sim$ 7 descendant to us would have a B magnitude of ~ [$M_*$
  +(m-M) + 9 mag (dust)] $\sim$ - 20 + 29 + 9 $\sim$ 18 magnitude.
                  
Attempts have been made to find ultra-compact galaxies by setting
spectroscopic fibres on bright starlike objects superposed on clusters
(Phillipps et al, 1998, Drinkwater and Gregg 1998). There was some
limited success with the discovery of Ultra-Compact-Dwarf
galaxies. However we would expect that most, and certainly most of the
bolometrically luminous ones, will be choked in their own
smoke(dust). They might however turn up in dedicated searches in the
FIR.
                  
The above discussion is highly simplistic, but the conclusions are so
strong that one hardly needs to qualify them further. Even if they
survive intact around us today the descendants of redshift 7
Exponential galaxies would pass unnoticed without a dedicated and
extensive search in the FIR.
                  \section{ HOW ELLIPTICAL GALAXIES SINK}
                  
For simplicity we have so far concentrated exclusively on
Exponentials. We now turn to giant Ellipticals which have the
softer light distribution:

\be ln\left(\frac{I(\theta)}{I_0}\right)=-\left(\frac{\theta}{\alpha}\right)^{1/4}\ee
                  
though that also implies a small amount of luminosity in a sharp pip
in the core. At first sight it will look as if Ellipticals have very
different Visibility functions from Exponentials, reaching their peak
angular-size Visibility $V^\theta$at a central SB no less than 7 mag brighter than
Exponentials (Disney, 1976). However that turns out to be an artefact
of the parametrization, and if a more physical SB measure $\mu_{1/2}$ ( the SB at
half light radius) is introduced then one finds [ e.g.Davies 1990] that the Elliptical
and Exponential Visibilities lie almost on top of one another, but
with the FWHM of the Ellipticals being somewhat broader (4.2 mag as
opposed to 2.5).
                  
The algebra is much the same with the following modifications:

                  Eqn (3):\be L(\infty)=8!\pi I_0\alpha ^2\ee
                  
                  In Eqn. (8):$\{  \}\rightarrow \{(0.92)^4 exp(-0.92\dmu/2)\}$
                  
                  where the maximum of$\{\}$ occurs at $\dmu$ = 4/0.46=8.7  mag.

                 Also in (8): \be\sqrt{\frac{2}{\pi}} \rightarrow \sqrt{\frac {2}{\pi 8!}}\ee
                 
                 $d^m$ is identical but with $f(\dmu)$  Eqn (15) replaced by $f_E (\dmu)$     where
                 
                 \be f_E (\dmu) \equiv \frac{L(\dmu)}{L(\infty)} = 1-e^{-y}( 1+y+ \frac {y^2}{2!} + ..........+ \frac{y^7}{7!})\ee
                 
                 with
                 
                 \be y=(\theta/ \alpha)^{1/4} = 0.92\dmu\ee

To plot the two Visibilities $V^\theta $ and $V^m$ together we need first to adopt a
value for $\Gamma_c$ , and as we shall be most interested in apparently small
distant galaxies we adopt a value at the upper limit of the $\Gamma_c$  range of $\Gamma_c$ =5 (
Eqn 21) .That leads to a crossover point P at a contrast  $\dmu(P)$ = 10.4 mag
and thus to the Visibility diagram shown in Fig 7
                 
\begin{figure*}
\begin{center}
\includegraphics[width=3in]{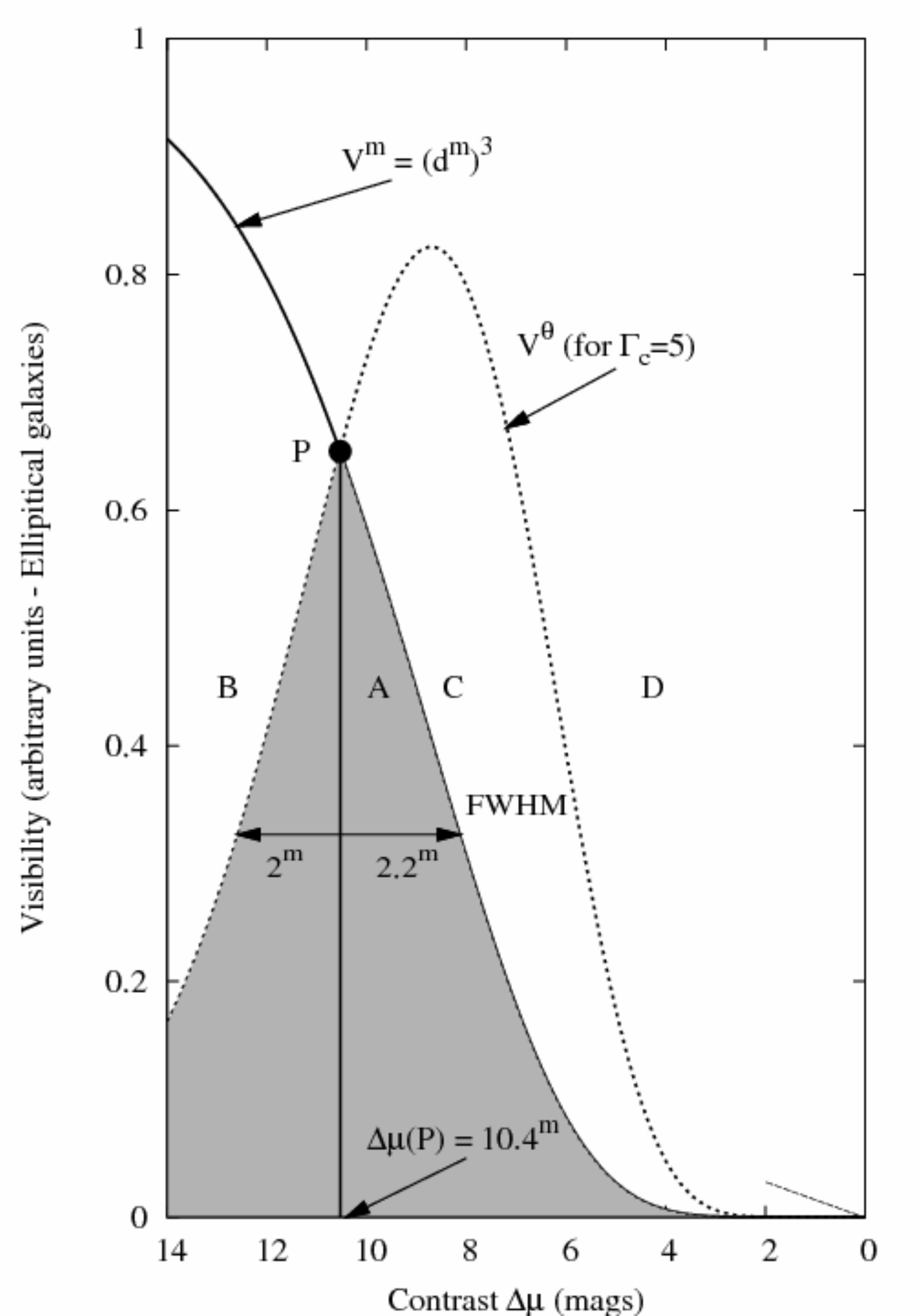}
\caption{Visibility as a function of surface-brightness-contrast
           $\dmu \equiv(\mu_c-\mu_0)$ for Giant Elliptical
           Galaxies. The magnitude-limited Visibility $V^m$ is
           normalised to 1 at  high contrast i.e. high SB towards
           the left because all the galaxy's light will show
           above the sky then. The angular-size Visibility $V^\theta$
           is the humped function . To the left its apparent size
           shrinks as, for a given luminosity, a galaxy must physically
           shrink as its SB increases. To the right it shrinks as
           more and more of its outer light is lost below the
           sky. The relative heights of $V^m$ and $V^\theta$ are
           determined by the pure number $\Gamma_c=l_c / I_c
           \theta_c^2$ which must have a value close to 5 for all but
           very nearby galaxies hundreds of pixels in diameter (Sect
           IV). The actual Visibility is the lower envelope of both
           curves i.e. the shaded area marked A with a peak at P
           where the two Visibilities intersect  to make a
           Wigwam. Only galaxies within A will be detected in a
           survey with limits $(l_c,\mu_c,\theta_c)$ in the combination
           $\Gamma_c$ as above . Galaxies in B will appear too small
           to be distinguishable as such, those in C too faint, those
           in D both too small and too faint. The FWHM of the
           Visibility window A is 4.2 mag, as opposed to 2.5  for Exponentials (see Fig 3). The
           location of the Elliptical peak P is at a contrast
           of $\dmu(P)$ = 10.4 mag, far higher than the Exponential
           peak at $\dmu(P)$ = 3.5 because, by comparison, Elliptical
           light distributions rise towards a sharper peak towards
           the core. However that peak contains very little light and
           so a fairer measure of the SB of a galaxy is the SB at
           half light = $\mu_{1/2}$, and a fairer comparison of the
           Visibilities of the different kinds of galaxies is made
           using their $\mu_{1/2}$s to measure a contrast
           `$\dmu_{1/2} \equiv(\mu_c - \mu_{1/2})$ , see Fig 9. [NB
           galaxies hundreds of pixels across have
           lower $\Gamma_c$ s, and so their $V^\theta$ falls below
           their $V_m$ , and so their Visibility is defined
           entirely by their $V^\theta$ ,Disney (1976) , and
           they don't have a Wigwam diagram.] }
\label{fig7}
\end{center}
\end{figure*}     
                  
Once again notice the two Visibilities intersect (at P) and the result
can only be another discontinuous, sharply peaked Wigwam Visibility
function limited on the right by  $V^m$and on the left by $V^\theta$ . The FWHM of the
combined Visibility curve enclosing the Visibility Window A is 4.2
mag, 1.7 mag on the Low SB side, and 2.5 mag on the High. The
normalisation shown is such as to make $V^m\rightarrow 1$ as the Contrast $\dmu\rightarrow\infty$ . The
Visibility shape, and some of the subsequent consequences, are
different from Phillipps et al (1990) because there no cognisance was
taken of $\Gamma_c$ , and so there were was no unambiguous way to adjust the
relative heights of $V^\theta $ and $V^m$ .
                  
Exactly as for Exponentials, Tolman dimming and Cosmology can be added
to yield the apparent magnitude $m_E(z)$ and angular-size (diameter) $\theta_E " (z)$  as (see Eqn (44):
                  
\begin{eqnarray} m_E(z)=M+25 -2.5 log f_E(\dmu') +5 log(c/H_0)\nonumber\\  + 5log [(1+z)\cdot ln(1+z)]\end{eqnarray}
                  
                  where $\dmu'=\dmu-10log(1+z)$

                  while (47):
\\
                  
\begin{eqnarray} \theta_E  " =\frac{4}{\sqrt{4\pi\cdot 8!}}\cdot \{(0.92\dmu')^4 exp (-0.92\dmu' /2)\} \nonumber\\ \times (1+z)\cdot \frac {H_0/c}{ln(1+z)}  \cdot dex(-5)\cdot dex(0.2(\dmu_c)\cdot dex(-0.2M) \end{eqnarray}

[The dex(-5) scale-factor accounts for the difference between the 10
  pc in (m-M) and the Mpc used in $H_0$ .]
                  
Fig 8 shows how the apparent magnitudes and sizes of giant Ellipticals
fade with redshift. It is the analogue to fig 6 for Ellipticals, and
like that diagram it too compares a galaxy of 'normal, i.e. 'local' SB with a
Masquerade, that is to say one which has a SB 9 mag more brilliant so as to
give a maximum Visibility at a z of 7. It is interesting to compare
Figs 8 and 6:
\begin{figure*}
\begin{center}
\includegraphics[width=5in]{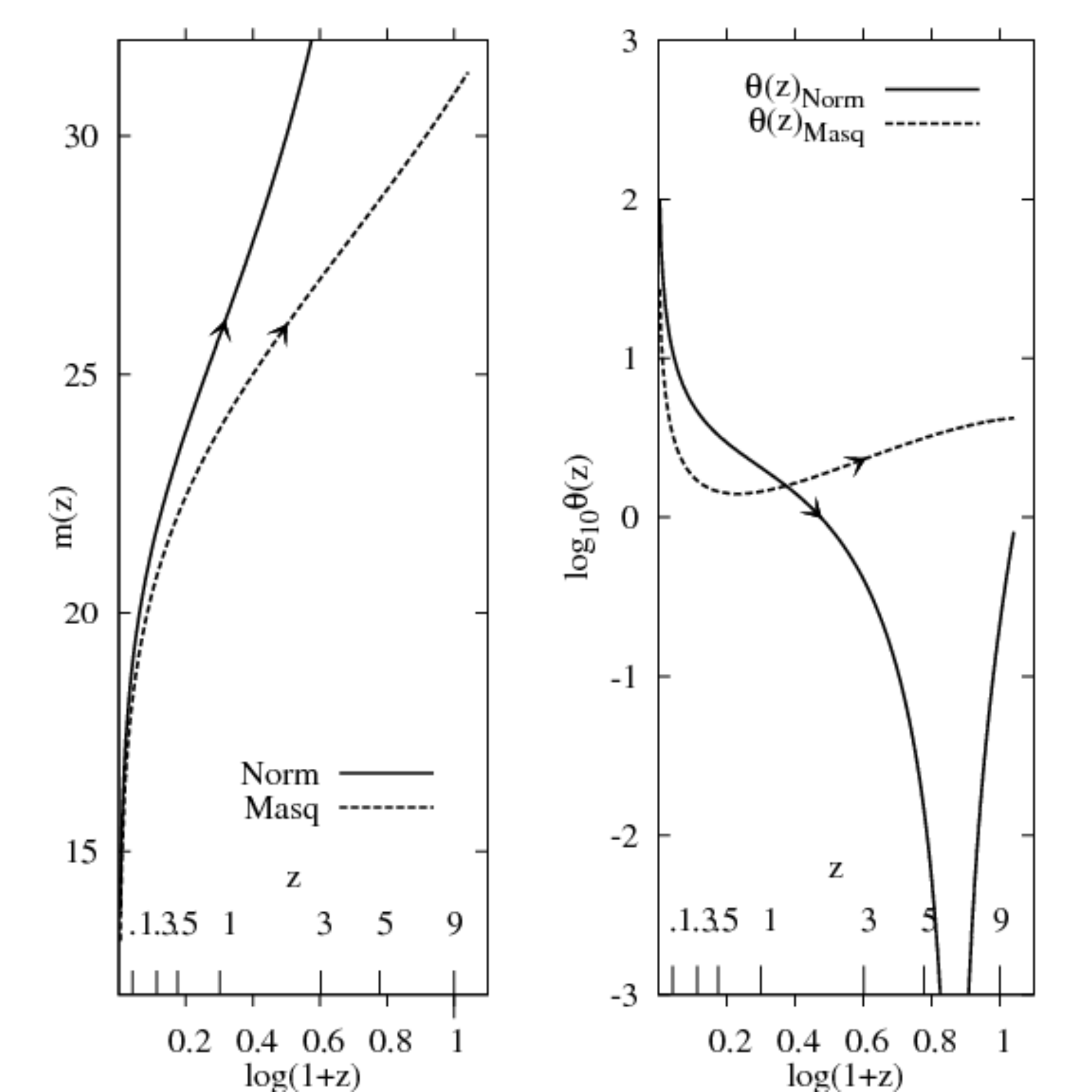}
\caption{The apparent magnitude $m(z)$ (left panel) and angular size
         (right) (diameter in arc sec) of an L* Elliptical with
         normal SB [$\mu_0\sim 15\mu,\mu_{1/2} \sim 23\mu$ in
         V band] and a Masquerade L* Elliptical with a SB 9 mag more
         brilliant, as a function of redshift z. The normal gE
         crosses the Hubble Deep Field line ( $\sim$ 30 mag) at z
         $\sim$2 but the Masquerade reaches $z\sim\ 5-6$ before it is
         extinguished , because it loses very little outer light
         below the sky. For angular sizes, aberration kicks in so the
         Masquerade reaches a minimum angular size of $\sim 2$ arc
         sec at $z \sim 1$ and it apparently grows gradually
         thereafter. [The odd Aberration effect seen on $\theta(z)$
         for the Normal gE is of no consequence, because by then its
         $m(z)$ has long since fallen below the sky.] Compare with Fig
         6 for Exponentials (see text). For a given Luminosity
         Ellipticals can be seen significantly further away than
         Exponentials, which might give the false impression that
         they formed earlier.}
\label{fig8}
\end{center}
\end{figure*}                   
                   
First notice the gentler slopes of m(z) for the EÕs at high z. They
do not fade away so quickly with z because they have a light
distribution with a steep core, and this reflects in a difference
between $f(\dmu)$ (Eqn 15) and $f_E(\dmu)$ (Eqn 61). So while a Milky Way dies completely,
both in size and magnitude at z =1.2, m(z) for a Giant gE doesnÕt
reach the typical HST Deep Field limit of $\sim$29 until z $\sim$2. This extra
Visibility range in $V^m$ allows the Aberration to kick in more decisively
ensuring the much more upturned $\theta_E (z)$ curves at high z. Thus the Normal gE
would appear large enough to be seen as a galaxy with HST out to a
redshift of 4-5 if, by z = 2, it hadnÕt already faded in magnitude
below 29.

Comparing Fig(8b) with (8a), the Masquerade Elliptical (i.e. the one with a SB 9 mag more
brilliant than a Normal gE ) is too compact to lose much of its outer
skirt of light below the sky so that it is only dimmed by distance,
Cosmology, and the Tolman $(1+z)^{-2}$ effect. The magnitude-difference between
Normal and Masquerade is entirely due to the aforesaid
skirt-effectÕ.
                    
                    In summary, Normal Giant Ellipticals ought to be seen out to higher zÕs (~2 with HST) than discs of the same Luminosity (z=1.2). This is the opposite conclusion to that reached by Phillipps et al (1990) and is accounted for purely by $\Gamma_c$ (Sect 4). This could give the misleading impression that gEÕs formed before other galaxies.
                    
The one dramatic difference between gEÕs and ExponentialÕs is
the position of $\dmu(P)$ at the point where the Visibility reaches a maximum,
i.e. at the centre of the Visibility Window A;  $\dmu(P)$ = 10.4 for Ellipticals there
whereas for Exponentials it = 3.5 . But this is an
artefact of the parameterisation. Ellipticals have a pip of light in
their core which yields a correspondingly bright $\mu_0$ which however is
representative of very little luminosity in total. Better therefore to
use $\dmu_{1/2}$ where :

\be \dmu_{1/2}  \equiv  \mu_c - \mu_{1/2}\ee

and  $\mu_{1/2}$ is the SB $\emph {at}$  the half-light radius.  

It is trivial to show:
                    
               \be (\mu_{1/2} - \mu_0)_{Exp} = 1.8\ee

                    Thus  \be \dmu_{1/2} (P) = 3.5 -1.81 =1.7\ee

  for Exponentials  while  for Ellipticals :

                    \be (\mu_{1/2} -\mu_0)_E = 8.4mag\ee

                    Thus  \be \dmu_{1/2} (P) = 10.4 - 8.4 \approx 2.0 \ee

   So in terms of a more representative
measure of SB, the half-light SB $\mu_{1/2}$  , the Visibilities of both breeds of
galaxies lie almost exactly on top of one another, see Fig 9. It
is a remarkable diagram that ought to give galaxy astronomers 
food for thought and we discuss it further in Sect IX. For now, at
least, it argues that both extreme light distributions lie so close
to one another, as far as Visibility is concerned, that so should all
the intermediate types.
                    
                    \begin{figure*}
\begin{center}
\includegraphics[width=4in]{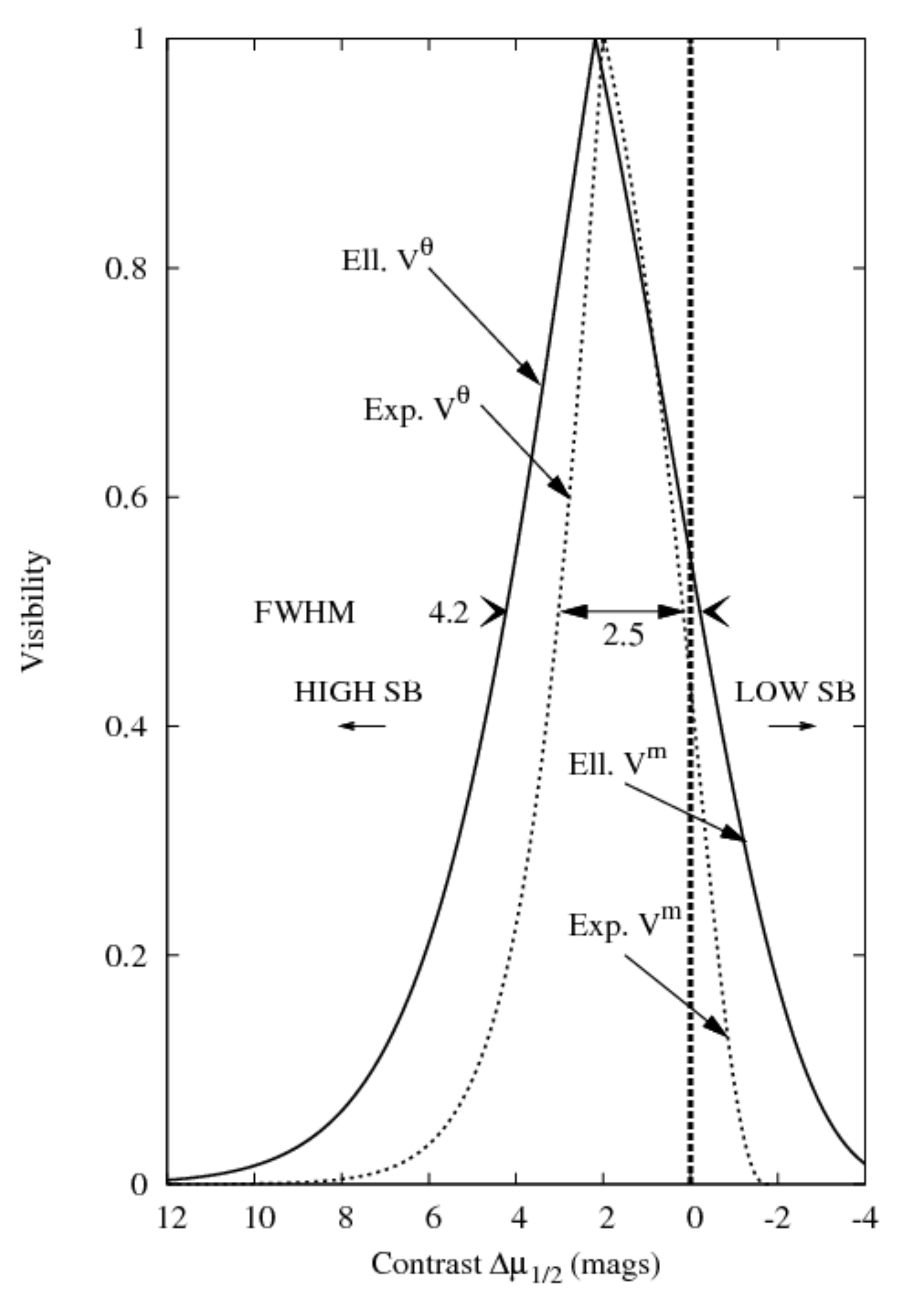}
\caption{ The Visibilities of both extreme morphologies of galaxy as a
  function of their contrast $\dmu$ with the sky, this time expressed
  in terms of their half-light surface brightnesses $\mu_{1/2}$ (Exponentials dotted, Ellipticals solid)
  . Now $\dmu\rightarrow\dmu_{1/2}\equiv(\mu_c -\mu_{1/2)})$ . In
  this more physically representative measure the two Visibility
  Wigwams fall almost on top of one another. They continue to the
  right of $\dmu_{1/2} = 0$ simply because half of their light lies
  below their $\mu_{1/2}$. What is most remarkable is how perilously
  close both Wigwams huddle to the sky. Galaxies that are marginally
  dimmer, for instance Dwarfs, or higher z objects, will quickly
  disappear altogether. That suggests a natural explanation for
  downsizing Ð as a selection effect which has nothing to do
  with evolution. [And what about the so called Missing Dwarfs
    in the CDM paradigm?]}
\label{fig9}
\end{center}
\end{figure*} 
     
\section{DOWNSIZING; A DIFFERENT EXPLANATION}
                     
Another remarkable phenomenon among apparently faint galaxies is
downsizingÕ [Cowie et al (1996), Heavens et al, (2004), Thomas et
  al , (2005), Noeske et al (2007a, 2007b), Perez-Gonzalez et al
  (2008)]. In purely observational terms it is the appearance of lower
luminosity and dwarf galaxies in apparent-magnitude selected samples
only at lower redshifts ( z $\leq0.5$). And when they do appear they have
comparatively blue colours and strong emission lines. If interpreted
in terms of the ESD hypothesis it requires giant galaxies to form
their stars first and dwarfs last, the very reverse of expectations
based on Hierarchical Galaxy Formation, the fashionable cosmogonic
hypothesis.
                     
But downsizing, as we shall next argue, can also be an entirely
natural outcome of the alternative SPD scenario, where it has no
implications for the ordering of galaxy evolution. The only assumption
required is that lower luminosity galaxies have, in a statistical
sense, dimmer intrinsic surface brightnessÕs. Because of obvious
selection effects such an assumption is not easy to demonstrate
unequivocally, but most observations, as well as common sense, speak
in its favour. For instance observations of HI- selected samples ,
which are unaffected by optical selection-effects, certainly show
such a correlation [Garcia-Appadoo et al. 2009, Chang et al 2011] with
                     
\be SB\propto L^{1/3}\ee

implying global stellar-densities that are independent of Luminosity
L. The 1/3 then arises because the path length through a more luminous
galaxy scales in that proportion.
                     
The rationale for downsizing under the SPDH is immediately apparent in
Fig 9. What we see there is that visible galaxies of all types ought
to huddle perilously close to the limit set by the local sky. And
observations going back to Holmberg (1965 ), Freeman (1970), Disney
(1976) and Davies et al (1994) testify that that this is  indeed the case
observationally. If now dwarf galaxies carry the further handicap of a
lower intrinsic surface brightness then they will naturally be the
first to sink below the sky. A smaller amount of redshift-dimming
should suffice to sink dwarfs entirely out of sight whereas giants
will still be visible further out. And, observationally speaking, that
is down-sizing.
                     
To see how potent this kind of downsizing is we calculate how rapidly
the Visibility of an Exponential galaxy (most dwarfs are Exponentials) will fall if we lower its SB
according to (69) and then redshift it. If we start with an
Exponential of optimum SB (i.e at the peak of the Visibility Window
with $\dmu$ = 3.5 mag) lowering its SB will run it towards the right in Figs,
1, 3 and 9 i.e. towards the boundary determined by $d^m$ and hence $V^m$ . In
those $V^m$ expressions the intrinsic SB enters explicitly only through $f(\dmu)$ , or
if redshift- dimming is allowed for in addition then through $f(\dmu')$ where as
usual $\dmu' =\dmu -10 log(1+z)$ . We can thus estimate what happens to the Visibility of a lower luminosity galaxy purely as a result of its extra SB dimming, by
comparison with an  $L_*$ galaxy of normal, i.e. local SB. Both galaxies will
of course fade with redshift but the dimmer dwarf will fade by more as
it is quickly swallowed up by the sky. The situation is best
summarised in the following Table 2 for 3 galaxies, an $L_*$ , an $0.1L_*$ and an
$0.01 L_*$, all obeying (69) and with the $L_*$ having optimum SB at the apex of
the Visibility Window at z=0.
 \begin{table*}
 \centering
 \begin{minipage}{70mm}
  \caption{ DIMMING AND DOWNSIZING}
  
  \begin{tabular}{@{}llllll|@{}}
  \hline
   $z$    & $L/L_*$  & $ \dmu $ & $\dmu' $ & $f^{3/2} $
   \\
   \hline
   0&1&3.5&3.5&0.76
   \\
   \hline
   0.1&& &3.1&0.69
   \\
   \hline
   0.3&&&2.4&0.52
   \\
   \hline
   0.4&&&2.0&0.41
   \\
   \hline
   0.5&&&1.7&0.31
   \\
   \hline
   0.6&&&1.5&0.25
   \\
   \hline
   0.7&&&1.2&0.17
   \\
   \hline
   0.8&&&1.0&0.11
   \\
   \hline1.0&&&0.5&0.02
   \\
   \\
   \hline
   $z$    & $L/L_*$  & $ \dmu $ & $\dmu' $ & $f^{3/2} $
   \\
   \hline 
   0&0.1&2.7&2.7&0.6
   \\
   \hline
   0.2&&&2.3&0.5
   \\
   \hline
   0.3&&&1.9&0.38
   \\
   \hline
   0.4&&&1.&0.17
   \\
   \hline
   0.5&&&0.9&0.09
   \\
   \hline
   0.6&&&0.7&0.05
   \\
   \hline
   0.7&&&0.4&0.01
   \\
   \hline
   0.8&&&0.2&0.002
   \\
   
   \\
   \hline
   $z$    & $L/L_*$  & $ \dmu $ & $\dmu' $ & $f^{3/2} $
   \\
   \hline
   0.1&0.001&1.8&1.8&0.35
   \\
   \hline
   0.1&&&1.4&0.2
   \\
   \hline
   0.2&&&1.0&0.11
   \\
   \hline
   0.3&&&0.65&0.04
   \\
   \hline
   0.4&&&0.3&0.005
   \\

\end{tabular}
\end{minipage}
\end{table*}

                     TABLE 2
                     
Recall that $f$ is the fraction of a galaxyÕs light still visible above
sky, and that $f^{3/2}$ is proportional to the Visibility . The Giant doesnÕt
sink completely until z $\approx $1.2, the Low Luminosity galaxy has virtually
gone by z=0.7 and the Dwarf by 0.4.
                     
The other important aspect of downsizing as observed is that the low
luminosity and dwarf galaxies are bluer and have stronger emission
lines at a given redshift. But that could be explained by a natural
SB-selection-effect. Galaxies approaching total immersion would be far
more prominent if they were undergoing temporary bursts of star-formation,
with a consequent increase in SB. Take the dwarf in the table at
z=0.4. If its SB were to increase by 0.35 mag we can see that its
Visibility would increase by a factor (0.04/0.005) or 8, and if by 0.7
mag then by over 20. (The 'Half-bakedÕ' appearances of many HDF
galaxies could be due to the extra star-formation required to lift
such galaxies above the sky.)
                          
In summary then the SPD hypothesis has a natural explanation for the
downsizing observations in terms of Visibility. It assumes only that
intrinsic SBs generally fall with Luminosity, and implies nothing
about Evolution.
                          
[PS: One wonders if a similar effect could explain the so called
 ÔMISSING DWARFSÕ problem which afflicts CDM (e.g. Klypin et al
  1999 ). If the correlation between SB and Luminosity in Eqn (69)
  holds all the way down to very faint objects then their SBs and
  Visibilities would render them exceedingly hard to find, even
  nearby. Eqn (69) implies that  $\dmu_{1/2}=-1/3\cdot \Delta M$. Thus a .001 L* Exponential Dwarf would
  have a SB= $-1/3\times 7.5=2.5$ mag dimmer than a giant at the peak of the Window, and its
  Visibility would consequently be very small. And if the correlation
  were slightly steeper,e.g.$ \Delta \mu_{1/2} =-1/2\cdot\Delta M$ , which is probably not ruled out by the
  observations, then the $\Delta \mu(P)$ would be 3.5-(7.5/2), i.e. would be negative
  and such a  Peak dwarf would be totally and irretrievably sunk below the
  sky at $\emph {any}$ redshift.]
                          
                          \section{DISCUSSION}
                          
If the universe is expanding then the associated Tolman dimming should
render conventional galaxies undetectable at high z . Most should be heavily affected
 by the sky at redshift 0.5, all totally submerged by redshift 2. The fact that
we can easily see galaxies out to redshift 7 means either that
conventional galaxies have undergone the most dramatic evolution (The
Evolving Single Dynasty or ESD hypothesis), or that the galaxies out
there belong to different populations, different dynasties, whose
descendents havenÕt so far been identified nearby (The Succeeding
Prominent Dynasty Hypothesis or SPDH).
                          
If our neighbourhood galaxies are role models then high redshift
galaxies are truly bizarre. They are one or two orders of magnitude
smaller in physical size, while their intrinsic surface
brightnessÕs must be 9 mag or 4000 times higher. Moreover, and this
is even more extraordinary, they must have systematically adjusted
their sizes and their SBs over cosmic time so as to squeeze themselves
through the narrow Visibility Window at all the various intermediate
redshifts where they can be seen. Such dramatic evolution is hardly
consistent with the archaeology of nearby galaxies whose
star-formation-histories seem rather steady and quiescent over
time. Furthermore the high redshift galaxies are, in co-moving terms,
rather rare (e.g. McLure et al 2010), too few in number to provide the
ultraviolet radiation needed to re-ionize the IGM at redshifts between
6 and 11. And this difficulty is compounded if downsizing is a
physical, as opposed to an illusory phenomenon, for the lower
luminosity galaxies form too late to contribute to the ultraviolet
budget when it would be necessary.
                          
The SPD hypothesis requires no such dramatic evolution, and explains
both downsizing (Sect IX) and galaxy-expansion (Sect VI) as illusory
phenomena, the side effects of Visibility Theory,
i.e. SB-selection. It also leads, through its postulation of large
numbers of sunken galaxies, particularly at high redshift, to a
natural solution for the Re-ionization Problem. And it is interesting
that there are two other strong hints in the recent literature at the
presence of such a sunken high-redshift population: one due to what we
call ÔInfant MortalityÕ, the other to an excess of high-z
QSOALs. We discuss these next.
                          
ÔINFANT MORTALITYÕ.  In the ESD hypothesis the stellar mass
density at any epoch ought to equal the accumulated rate of star
formation over all preceding epochs:
                          
\begin{eqnarray}\rho(t_0)=\int _{0 } ^{t_0} \dot{\rho} (t)\cdot dt= \int _\infty ^{z_0} \dot {\rho} (z) \cdot \frac {dt}{dz}\cdot dz \nonumber\end{eqnarray}
                          
In the SPD scenario this however will no longer apparently be the case
because in moving from z to z+dz some lower SB galaxies will appear to
sink beneath the sky due to Tolman dimming $ (i.e.-(\partial \rho/\partial z)_{sink} \cdot \Delta z)$ while other higher-SB
objects will apparently surface (thanks to aberration) to partially
replace them .  Thus the integral above should be replaced by
                          
 \begin{eqnarray} \rho(t_0)=\;\int_{\infty}^{z_0}\,\left [ \dot{\rho}(z)\cdot \frac {dt}{dz} + \left\{\left( \frac{\partial \rho}{\partial z}\right)_{Surf} -\left ( \frac{\partial \rho}{\partial z}\right )_{Sink}\right \}\right] \cdot dz\nonumber \end{eqnarray}
                          
where the net $\{ \}$ could be either positive or negative, depending on
the distribution of galaxy numbers as a function of intrinsic SB. All
one can say for sure is that there is no reason to expect a good match
between observed stellar densities and accumulated past
star-formation. And such a mismatch has been noted by many
observers. For instance Perez-Gonzalez et al (2008), p 248, remark :
"We find that the cosmic SFR densities estimated by differentiating the
evolution of cosmic stellar mass density do not match the observations
based on direct SFR tracers as also noted by Rudnick et al (2006),
Hopkins and Beacom (2006) and Burch et al (2006). The mismatch up to
z $\sim\ $2 ( a factor of 1.7) could be explained by changing the IMFÉÉ
And as z rises from 2 to 4.5 the discrepancy is larger ( a factor 4 to
5 É..) "  This highly significant mismatch is in the sense that
less galaxies are observed in each redshift bin than the total of
previous SF would lead one to expect. Under the SPD hypothesis this
would be naturally  explained if most high redshift galaxies sink below the sky
once their most vigorous period of SF comes to an end. This is rather
direct evidence of the SPDH, though perhaps not conclusive.
                          
SUNKEN DLAs. Even where they cannot be seen in emission, sunken galaxies should
still show up in absorption, in particular as QSOALs, probably of the
Damped Lyman Alpha (DLA) i.e high-column-density variety. The number
detected in the redshift range :\\
                          
 \begin{eqnarray} N(z).dz \propto \int _{z} ^{z+\Delta z} \int _{0} ^{L} \phi (z,L)\cdot A(z, L).\cdot g(z)\cdot dz\cdot\nonumber dL\end{eqnarray}
 \\
                          
where g(z).dz is the physical path-length derived from some
cosmological model,  $\phi(z,L)$ is the co-moving density of galaxies of luminosity
L, and A(z, L) is their effective cross-section per galaxy. As is well
known (e.g. Wolfe et al 2005) if  $\phi(z,L)$  corresponds to the local value for  $L^*$ galaxies,
A(L,z) $\sim$ observed optical area of $L^*$s, and a reasonable extra boost
(by a factor 2-15) is allowed for the dwarf contribution associated
with the $L^*$s, then there is a fair correspondence with the observed N(z)
i.e. between 5 and 10 percent of high redshift QSOs ($z\leq 6$) have DLAs in
their spectra. Unfortunately that sets no absolute limit on $\phi$ , and
hence on a population of sunken galaxies, without some independent
knowledge of the cross -section A(L, z) Ð which is not
available. One can always increase $\phi$ by decreasing A. Nevertheless if
the SPD hypothesis is true one might expect a mismatch between the
z-dependence of N(z) and the number density inferred from the rate of
galaxy formation, in the sense that that there will probably be more
DLAÕs in the distance, corresponding to the extra galaxies out
there which have sunk from sight. And in a qualitative sense at least
that is what seems to be observed in the large SDSS sample of QSOÕs
recently analysed by Prochaska and Wolfe (2008). Instead of a decline
in cross section [our N(z), their l(z)] and in co-moving HI density
with rising redshift, caused by the decline in the number of
already-formed disc galaxies with z, they find instead an increase by
a factor of 2 between z=2 and z= 4.5 (only 2 Gyr.) which they find 
"a profound and surprising result". There may be other explanations
(which they mention) but it seems qualitatively consistent with the
idea of a larger proportion of sunken galaxy absorbers at higher
redshifts.
                          
So there is significant indirect evidence in favour of the SPD
hypothesis and its implication that the universe is stuffed with
Hidden Galaxies. Moreover the SPDH is an almost inevitable consequence
of Visibility Theory ,which is hardly radical, but usually
neglected. If Hidden galaxies are not ubiquitous it will take a great
number of fortuitous coincidences to explain why all the detected
galaxies in the universe have arranged themselves, at all redshifts,
so as to squeeze through our narrow, parochial, Visibility Window (e.g. Fig 2).

Indirect evidence is all very well but direct evidence of Hidden
Galaxies, particularly of such a rich population as the SPD hypothesis
requires, would be far more persuasive. If Hidden galaxies are so
common why havenÕt they turned up in dedicated searches with large
telescopes, and why havenÕt far more of them appeared in the blind
HI surveys (that are of course free of SB selection) that we and
others have recently been carrying out?
                          
With the benefit of hindsight we can answer both questions. The
argument at the end of Section IV is new. We and others understandably
supposed that a large enough optical telescope, fitted with CCDs and
dedicated to the search, would turn up Low SB galaxies , if they
exist. But alas that is simply not true , and Eqn (34) reveals
why. When it comes to searching for LSBGs Tolman dimming together with the small sizes of CCDs more
than cancel out their high quantum efficiencies and infinite dynamic
ranges. The Schmidt photographic surveys, completed in the 1980s,
represent the best that can be done. That is a depressing admission, but
there seems to be no practical way around Eqn.'s  (34) and (41) in Sect IV.
                          
The blind 21 cm surveys which we and others have carried out such as
HIPASS, (Meyer et al 2004), HIJASS(Lang et al 2003), AGES(Cortese et al 2010) and ALFALFA (Giovanelli et al. 2005) have turned up only one
truly invisible giant galaxy ,Virgo HI21 (Davies et al. 2004, Minchin et al, 2005, 2007),
and there is even some dispute about that ( Haynes et al 2007). In
fact HIPASS 's failure to turn up a single invisible galaxy among
its 4000 detections in the Southern hemisphere, has been used to claim
(Doyle M.T. et al 2005, Wong et al 2006, Wong et al 2009) that Hidden
galaxies do not exist, or are very rare (and we were party to that
claim). Our re-analysis (Disney 2008, Disney and Lang 2011) shows however that the
claim was based on a grossly optimistic estimate for the reliability
of the optical identifications involved.  When clustering is allowed
for there could well be $~ $100 dark galaxies in the sample i.e. HI
sources that have been misidentified with optically bright objects
that are fortuitously close by in both angular and redshift space. And blind
surveys with a larger dish won't improve the situation, because
their extra resolution is exactly counterbalanced by the extra
distance at which the typical sources will be found ( Disney
2008). Anyway when objects such as Malin 1 certainly exist, a giant
LSBG 200 kpc across, containing $ > 10^{11} $ solar masses of HI, (Bothun and Impey
1989) one has to be cautious about blind-scanning techniques in
general. Such objects nearby will be much larger than the scanning
radio beams, and so tend to be lost in the process of noise
subtraction. None of the existing blind HI surveys, in our opinion,
sets strong constraints on the presence of HI-rich Hidden Galaxies
nearby. In such surveys absence of evidence is not strong evidence of
absence.
                          
Nevertheless it remains vital to pin some of the hypothetical Hidden
Galaxies down. At low redshifts in the HI suspicious optical
identifications should be vigorously pursued with interferometers. And
the compact decendants of the z=7 galaxies, now small, faint and
choked in dust (Sect VII) might still be locatable in the FIR relatively nearby. At
high redshift sunken Milky Ways beyond a z of 1.2 may have regions, if
seen in emission lines, that will still rise above the sky in
objective prism surveys. Moreover they, and their Elliptical
counterparts (z $>$ 2) should still give rise to faint supernovae which
may turn up in what otherwise appears to be intergalactic space, and a start has been made in such surveys( Hayward et al 2005).

In one sense one must hope that the SPD hypothesis is wrong, for if it
is right then extra-galactic research is going to be so much
harder. The obvious program of decoding galaxy-formation and evolution
simply by building larger instruments such as JWST or ELT to look at fainter,
more distant objects wonÕt work because a given dynasty of galaxies
will remain visible through our Visibility Window for a only a limited
range of redshifts, i.e. for only a restricted portion of its life. We
might see the infants of one dynasty, the children of another, the
adults of a third, and the grizzled elders of a fourth only among our
neighbours.
                          
On the other hand the SPD hypothesis has strong epistemic
advocates. It is extremely parsimonious (Gauch 2005) relying as it
does only on Tolman dimming and Visibility Theory, the last of which
we have been at some pains to explain and defend. Neither is it the least
radical in the sense that it employs assumptions outside very ordinary
physics. And it is vulnerable in that it predicts the existence of
whole dynasties of galaxies which are presently undetected, but whose
existence it may eventually be possible to affirm or deny.
                          
\section{ACKNOWLEDGEMENTS}
                                                   
We thank Mathias Disney (UCL Geography) for the diagrams, Nino Disney, Richard Elliott,
 and Joe Romano (U Texas, Brownsville) for indispensable
help with word processing, and Peter Coles for
drawing our attention to Ned WrightÕs very useful ÔCosmology
CalculatorÕ. MJD would like to thank fellow members of the Hubble
Space Telescope FOC and WFC-3 camera teams from whom he learned so
much between 1976 and 2011, and the European Space Agency (via the
European Space Telescope Coordinating Facility) for travel support over that period.

\end{document}